\documentclass[11pt]{article}	
\usepackage{amsmath,amsthm,amsfonts,amscd, amsmath} 
\usepackage{amssymb,amsfonts,amsmath,amscd,amsthm}	
\usepackage{graphicx}
\usepackage{setspace}
\usepackage[caption=false]{subfig}  
\usepackage{verbatim}   	
\usepackage{makeidx}    	
\usepackage{color}
\usepackage{enumerate}
\usepackage{ upgreek }
\usepackage{authblk}
\usepackage{epsfig}     	
\usepackage{url}		
\usepackage{tcolorbox}
\usepackage{multirow}

\usepackage{booktabs} 

\graphicspath{{../figs/}} 


\theoremstyle{definition}

\theoremstyle{remark}

\usepackage[toc,page]{appendix}

\newcommand{\latexe}{{\LaTeX\kern.125em2%
           \lower.5ex\hbox{$\varepsilon$}}}

\chardef\bslash=`\\	

\makeatletter		
\def\square{\RIfM@\bgroup\else$\bgroup\aftergroup$\fi
 \vcenter{\hrule\hbox{\vrule\@height.6em\kern.6em\vrule}%
                       \hrule}\egroup}
                                               
\makeatother		
\makeindex  

\usepackage{fancyhdr}
\newcommand{\blue}[1]{\textcolor{black}{#1}}

\begin{document}
\title{
A Predictive Multiphase Model of Silica Aerogels for Building Envelope Insulations
}

\author[1]{Jingye Tan\footnote{These authors contributed equally to this work.}}
\author[2]{Pedram Maleki$^*$}
\author[1]{Lu An}
\author[1]{Massimigliano Di Luigi}
\author[3]{Umberto Villa}
\author[4]{Chi Zhou}
\author[1]{Shenqiang Ren}
\author[1]{Danial Faghihi\thanks{corresponding author, danialfa@buffalo.edu (D. Faghihi)}}
\affil[1]{Department of Mechanical and Aerospace Engineering, University at Buffalo}
\affil[2]{Department of Materials Science, Sharif University of Technology, IRAN}
\affil[3]{Electrical and Systems Engineering, Washington University in St. Louis}
\affil[4]{Department of Industrial and Systems Engineering, University at Buffalo}
\renewcommand\Authands{ and }

\maketitle


\begin{abstract}

\blue{
This work develops a systematic uncertainty quantification framework to assess the reliability of prediction delivered by physics-based material models in the presence of incomplete measurement data and modeling error. The framework consists of global sensitivity analysis, Bayesian inference, and forward propagation of uncertainty through the computational model.
The implementation of this framework on a new multiphase model of novel porous silica aerogel materials is demonstrated to predict the thermomechanical performances of a building envelope insulation component. 
The uncertainty analyses rely on sampling methods, including Markov-chain Monte Carlo and a mixed finite element solution of the multiphase model.
Notable features of this work are investigating a new noise model within the Bayesian inversion to prevent biased estimations and characterizing various sources of uncertainty, such as measurements variabilities, model inadequacy in capturing microstructural randomness, and modeling errors incurred by the theoretical model and numerical solutions.
}

\end{abstract}


\clearpage
\section{Introduction}


Commercial and residential buildings are responsible for nearly 40\% of energy consumption and greenhouse gas emissions in the United States, of which approximately 39\% are through the building envelopes such as the floor, roof, and walls \cite{USeng}. Novel insulation materials with ultralow thermal conductivity and sufficient mechanical stability are essential to reduce building energy consumption. Moreover, substantial heat flow sources through the building envelope are the thermal bridges at the interface of assemblies such as between windows and walls and parapet-wall-roof intersections, leading to heat flow bypassing the insulation \cite{alhawari2018}. Thermal breaks are the insulating components incorporated within the envelope to interrupt the heat flow path and reduce the undesired effects of thermal bridges \cite{topuzi2020, scott2019}. Silica aerogels' lightweight and ultrahigh insulation properties make them the most promising high-performance materials for the next-generation building insulations, e.g., \cite{gao2014, gao2014II, berardi2017benefits, cuce2014toward, jelle2015}. Silica aerogels consist of a mesoporous internal structure of silicon dioxide bounded chains with high porosity (around 90\%) and specific surface area that results in their superinsulation performance \cite{karamikamkar2020, maleki2014, feng2018, an2021reflective}. 
Low mechanical strength, long processing time, and small-dimension production related to the supercritical drying fabrication process are the main impediments preventing the widespread adoption of silica aerogel in the construction industry.
We have recently developed a new synthesis method that overcomes these challenges leading to the fabrication of mechanically strong and cost-effective silica aerogels \cite{yang2019hierarchical, an2020all, an2021wearable}. Contrary to supercritical drying, where the porous aerogel structures are produced by removing the solvent without collapsing the solid structure, the new method relies on an in-situ bubble-supported pore formation. Combined with ambient pressure and temperature drying, this approach reduces aerogel fabrication energy, time, and equipment cost \blue{by approximately 60\%}.
Relying on the new synthesis, we have demonstrated additive manufacturing of aerogel components using direct ink writing (DIW) \cite{chi2021}. The DIW additive manufacturing of silica aerogel enables printing non-structural thermal breaks with customized geometry compatible with the envelope architecture \blue{(see Figure \ref{fig:aerogel}(a))}. 
However, to bridge the gap between laboratory discovery \blue{and} their deployment in building applications, there is a well-recognized need for computational models to predict the new silica aerogel performance during building operations.


Numerous physics-based computational models, based on continuum descriptions of the materials, can be \blue{constructed} to simulate the thermal and mechanical responses of porous materials in practical time scale and domain size. 
A category of these models is based on homogenization theory, which assigns effective physical properties to the heterogeneous material, e.g., \cite{ torquato2002}. The recent developments of homogenization models have enabled accounting for the microstructural mechanisms, nonlocality, and multiscale responses of a wide range of materials, e.g., \cite{kim2020, matouvs2017review, oskay2020discrete}. 
Another class of computational models of porous materials is based on continuum mixture theory, which provides a general framework for modeling the action and interaction of multiple solid and fluid phases, e.g., \cite{rajagopal1995, bowen1980}. The fundamental idea underlying the mixture theory of materials is that each phase follows its own motion, balance laws, internal energy, and entropy while conforming with the fundamental physical laws of the mixture \cite{jha2020bayesian, OdenHawkins2010, faghihi2020, patki2020}. 
\blue{Different modeling assumptions result in a large number of possible models with different fidelity and complexity for a material undergoing particular physical processes.
It is imperative to characterize uncertainty in these models to guide choosing valid models for simulating the complex material behavior.
}


While there have been extensive computational studies of thermal and mechanical properties of porous materials (including aerogel), e.g., \cite{wang2019simulation, yang2018, gandomkar2018}, assessing the credibility of these models in predicting the material responses beyond observational data remains a challenging problem.
\textit{Physics-based predictive modeling} is a notion that has emerged in recent literature to describe the systematic use of data to dramatically enhance the power of computational models to predict the state of physical systems in the absence of direct observation \cite{odenbabuska2017, oden2010computer}.
Predictive computational modeling of materials systems entails ``training'' (i.e., calibrating) the material model using experimental or simulation data and quantifying the uncertainty in the predictions delivered by these models.
Sources of uncertainty in computational modeling of porous materials include limited observational data (specifically for relatively new materials such as silica aerogels), microstructural randomness, and inadequacy of the computational model in depicting complex physical mechanisms governing the material behavior.
This gives rise to the need for uncertainty quantification (UQ) methods to characterize the variabilities in experimental and randomness in model parameters, as well as propagation of uncertainty through the model to the prediction of quantities of interests (QoIs) \cite{ding2021, song2019, tan2021}. Coping with these uncertainties and assessing the reliability of computational prediction are the most crucial scientific obstacles in computational modeling and design of materials systems \cite{McDowell2020, panchal2013}.


\blue{
This contribution presents a general computational framework to systematically quantify uncertainty in physics-based models using incomplete measurement data and characterize the reliability in computational prediction of complex physical systems in the absence of observational data. 
The framework combines three uncertainty assessment methods:
First, the variance-based global sensitivity analysis characterizes the effect of parametric uncertainty on the model prediction and guides the reducing dimension of the uncertain parameter space.
Second, Bayesian inference enables learning the probability distribution of model parameters from incomplete and possibly noisy experimental data while accounting for modeling errors.
Finally, the solution of the statistical forward problem assesses the reliability of calibrated physics-based model prediction.
The novel aspects of Bayesian calibration in this work are 
implementing a new noise model that leverages hyperparameters to characterize uncertainty in data and modeling error
 along with the construction of informative priors using sensitivity analysis.
This framework is applied to a multiphase model of novel insulation silica aerogel materials, and numerical predictions of the thermomechanical performances of a building envelope thermal break are demonstrated.
To this end, a new thermodynamically consistent coupled heat transfer and elastic deformation model of silica aerogels is developed based on the continuum theory of mixture. The model accounts for solid (aerogel structure) and fluid (pores) phases and their interactions to represent the material response. A mixed finite element formulation is used to solve the system of equations governing the multiphase model. 
The uncertainty {quantification (UQ)} framework is then implemented on the multiphase model using sampling methods to accurately explore the parameter posteriors and characterize their interactions, given limited thermal and mechanical {experimental} measurements of silica aerogel materials.
Leveraging this UQ framework and corresponding computational implementation, various sources of uncertainty in computational prediction are characterized, including uncertainty in measurements, the inadequacy of the model to capture microstructural randomness, and
modeling errors incurred by simplifying assumptions in the theoretical model and numerical solutions.
}


The rest of this manuscript is organized as follows. Section 2 presents a brief derivation of the thermomechanical two-phase model based on the mixture theory. The UQ approaches, including parameter sensitivity analyses, Bayesian inference, and forward uncertainty propagation for predicting QoIs, along with their numerical solutions, are described in Section 3. The experimental data, finite element solution of the multiphase model, and uncertainty analyses and model prediction results are presented in Section 4. Finally, the discussion and conclusions are given in Section 5.

\section{Thermo-mechanical Theory of Multiphase Mixtures}\label{sec:mixture}

We propose a theoretical model of silica aerogel's thermomechanical behavior founded in the continuum theory of mixtures \cite{truesdell1962, bowen1980, eringen1965}. 
This section summarizes a general theory governing a continuum mixture of \blue{multi-}constituents, coupling heat transfer, and elastic deformation \blue{and then derives the two-phase model of porous aerogel materials}.
The mixture theory assumes that the reference configuration consists of $\blue{M}$ constituent species simultaneously occupying the same physical space.
The body undergoes a motion which maps the reference configuration onto a current configuration, with the spatial position of material points $\mathbf{x}$ at time $t$ defined by
\begin{equation}\label{eq:motion}
\mathbf{x} = \mathcal{X}_{\alpha} (\mathbf{X}_{\alpha},t),
\end{equation}
where, $\mathcal{X}_{\alpha}$ is the motion and $\mathbf{X}_{\alpha}$ is the material point positions of the $\alpha$-th constituent ($\alpha=1, 2, \cdots, M$).
The volume fraction of the $\alpha$-th constituent are defined by\footnote{Throughout this section  
we use the abbreviated notation $\sum_{\alpha}=\sum_{\alpha=1}^{M}$.
}
\begin{equation}\label{eq:volfrac}
\phi_{\alpha}(\mathbf{x},t) = \frac{dV_{\alpha}}{d V}
\qquad {\rm with} \qquad
\sum_{\alpha} \phi_{\alpha}=1,
\end{equation}
where $dV$ is a differential volume containing the point $\mathbf{x}$, and
$dV_{\alpha}$ is the proportion of volume occupied by constituent $\alpha$.
In addition, the velocity $\mathbf{v}_{\alpha}$ and
symmetric part of the velocity gradient $\mathbf{D}_{\alpha}$ of each constituent are defined respectively by
\begin{equation}\label{eq:velocity}
\mathbf{v}_{\alpha}(\mathbf{x},t)=\frac{\partial \mathcal{X}_{\alpha}(\mathbf{X}_{\alpha},t)}{\partial t},
\qquad
\mathbf{D}_{\alpha}=\frac{1}{2} (\nabla \mathbf{v}_{\alpha} + \nabla \mathbf{v}_{\alpha}^T).
\end{equation}
Correspondingly, the mixture velocity is,
\begin{equation}
\mathbf{v} = \frac{1}{\rho} \sum_{\alpha} \rho_\alpha \phi_\alpha \mathbf{v}_{\alpha},
\end{equation}
where $\rho_\alpha$ is the \blue{mass} density of each constituent and $\rho$ is the mass density of the mixture \cite{deboer2012, rajagopal1995}.
Finally, the relation among Lagrangian and Eulerian time derivatives are
\begin{equation}\label{eq:timederv}
\frac{d^\alpha\varphi}{dt} = \frac{\partial\varphi}{\partial t} + \mathbf{v}_\alpha \cdot \nabla\varphi,
\end{equation}
where ${d^\alpha\varphi}/{dt}$ is the material time-derivative related to the motion of each constituent and 
$\varphi(\mathbf{x},t)$ is any differentiable function.

%

\subsection{The balance laws for mixture}

The thermo-mechanical mixture theory follows the balance laws for mass, momentum, energy, and the inequality for entropy.
These basic laws are the essential building blocks on which to develop a physics-based material model.
According to the mixture theory, each constituent must satisfy its individual balance laws.

\textit{Balance of mass:}
\begin{equation}\label{eq:mb}
\frac{\partial (\rho_\alpha \phi_\alpha)}{\partial t} + \nabla \cdot (\rho_\alpha \phi_\alpha \mathbf{v}_\alpha) = S_\alpha.
\end{equation}

\textit{Balance of linear momentum:}
\begin{equation}
\rho_\alpha \phi_\alpha \frac{d^\alpha \mathbf{v}_\alpha}{dt} = \nabla\cdot \mathbf{T}_\alpha + \rho_\alpha \phi_\alpha \mathbf{b}_\alpha + \mathbf{m}_\alpha.
\end{equation}

\textit{Balance of angular momentum:}
\begin{equation}
\mathbf{M}_\alpha = \mathbf{T}_\alpha - \mathbf{T}_\alpha^T.
\end{equation}

\textit{Balance of energy (first law of thermodynamics):}
\begin{equation}\label{eq:energy_balance}
\rho_\alpha \phi_\alpha \frac{d^\alpha \varepsilon_\alpha}{dt} = 
\mathbf{T}_\alpha : \mathbf{D}_\alpha - \nabla \cdot \mathbf{q}_\alpha + \rho_\alpha \phi_\alpha r_\alpha + e_\alpha.
\end{equation}
In the above relations, 
$\mathbf{T}_\alpha$ is the partial Cauchy stress tensor,
$\mathbf{b}_\alpha$ is the body force per unit mass, 
$\mathbf{M}_\alpha$ is the intrinsic moment of momentum,
$\mathbf{q}_\alpha$ is the partial heat flux,
and
$r_\alpha$ is the external heat supply,
associated to the $\alpha$-th constituent.
Since we do not account for the electromagnetic effects, we restrict our attention to nonpolar materials $\mathbf{M}_\alpha=0$, in which the partial stress tensors are symmetric. 
Additionally, the interactions among constituents are characterized by
$S_\alpha$, $\mathbf{m}_\alpha$, and $e_\alpha$
that are the mass, momentum, and energy supplied to the constituent $\alpha$ by other constituents.

The balance laws for the mixture are defined by
\begin{eqnarray}
\frac{\partial \rho}{\partial t} 	&+& \nabla \cdot (\rho \mathbf{v}) = 0,\\  
\frac{\rho d\mathbf{v}}{dt} 	&=& \nabla \cdot \mathbf{T} + \rho \mathbf{b},\\
\rho \frac{d \varepsilon}{dt} 	&= &
\mathbf{T} : \mathbf{D} - \nabla \cdot \mathbf{q} + \rho {r},
\end{eqnarray}
where $\rho$, $\mathbf{T}$, $\mathbf{b}$, $\varepsilon$, ${r}$, and $\mathbf{q}$ are 
the mass density, the stress, the body force per unit mass, the internal energy per unit mass, heat supply, and the heat flux for the mixture.
The sum of the constituent balance laws over all constituents is compatible with the above equations if \cite{bowen1980, deboer2012}
\begin{eqnarray}\label{eq:mix_sum}
\mathbf{T} = \sum_\alpha \left( \mathbf{T}_\alpha - \rho_\alpha\phi_\alpha \mathbf{p}_\alpha \otimes \mathbf{p}_\alpha \right)
,\quad 
\mathbf{b} = \frac{1}{\rho} \sum_\alpha\rho_\alpha\phi_\alpha \mathbf{b}_\alpha, \\
\mathbf{q} = \sum_\alpha \left(\mathbf{q}_\alpha - \mathbf{T}_\alpha^T \mathbf{p}_\alpha + 
 \rho_\alpha\phi_\alpha \varepsilon_\alpha \mathbf{p}_\alpha + \frac{1}{2} \rho_\alpha\phi_\alpha \mathbf{p}_\alpha (\mathbf{p}_\alpha \cdot \mathbf{p}_\alpha) \right)
,\quad
 \mathbf{r} = \frac{1}{\rho}\sum_\alpha \rho_\alpha\phi_\alpha \mathbf{r}_\alpha \nonumber,
\end{eqnarray}
where $\mathbf{p}_\alpha = \mathbf{v}_{\alpha} - \mathbf{v}$ is the diffusion velocity of each constituent and the relation below holds for the mixture,
\begin{equation}
 \sum_{\alpha} \rho_{\alpha} \phi_{\alpha} \mathbf{p}_{\alpha} = 0.
\end{equation}
Additionally, for any mixture, the mass, momentum, and energy supplied to the $\alpha$-th constituent must satisfy the following conditions \cite{OdenHawkins2010},
\begin{equation}
\sum_\alpha S_\alpha = 
\sum_\alpha (S_\alpha\mathbf{p}_\alpha - \mathbf{m}_\alpha) =
\sum_\alpha \left[ e_\alpha + \mathbf{m}_\alpha \cdot \mathbf{v}_\alpha + S_\alpha (\varepsilon_\alpha + \frac{1}{2} \mathbf{v}_\alpha \cdot \mathbf{v}_\alpha) \right]
= 0.
\end{equation}
%

\subsection{Two-phase model of silica aerogel}

To simulate the thermo-mechanical behavior of silica aerogel,
we consider a binary mixture ($\blue{M}=2$) consisting of solid $\alpha=s$ and fluid (gas) $\alpha=f$ phases.
Assuming that the \blue{solid constituent of aerogel} is incompressible 
($\rho_s=$ constant), the mass balance (\ref{eq:mb}) for the solid phase is simplified to
\begin{eqnarray}
\frac{\partial \phi_s}{\partial t} + \phi_s \nabla \cdot \mathbf{v}_s = 0.
\end{eqnarray}
Substitution of the above relation into the fluid mass balance leads to the \textit{equation of state}, c.f. \cite{prevost1985wave},
\begin{equation}\label{eq:gen_eq_state}
\phi_f \nabla \cdot \mathbf{v}_f + (1-\phi_f) \nabla\cdot \mathbf{v}_s = -\frac{\phi_f}{\rho_f} \frac{\partial \rho_f}{\partial t}.
\end{equation}
Additionally, assuming no average shear viscosity in the fluid phase of aerogel, the stress in fluid and solid phases are obtained as, c.f. \cite{rajagopal1995},
\begin{eqnarray}
\mathbf{T}_f &=& -\phi_f p \mathbf{I} \label{eq:fluid_stress}\\
\mathbf{T}_s &=& -\phi_s p \mathbf{I} + \mathbf{T}'_s, \label{eq:solid_stress}
\end{eqnarray}
where $p$ is the \blue{intrinsic Cauchy} fluid pressure and $\mathbf{T}'_s$ is the effective solid stress.

Furthermore, defining the specific heat capacity at constant volume as $c_\alpha = {\partial \varepsilon_\alpha}/{\partial \theta_\alpha}$ and replacing the material derivative in (\ref{eq:energy_balance}) with a partial derivative according to (\ref{eq:timederv}), we obtain \cite{frey1996}
\begin{equation}
\rho_\alpha \phi_\alpha c_\alpha \left[ \frac{\partial \theta_\alpha}{\partial t} + (\mathbf{v}_\alpha \cdot \nabla \theta_\alpha) \right]= 
\mathbf{T}_\alpha : \mathbf{D}_\alpha - \nabla \cdot \mathbf{q}_\alpha + \rho_\alpha \phi_\alpha \mathbf{r}_\alpha + \mathbf{e}_\alpha,
\quad \alpha = s, f.
\end{equation}
%

The conservation laws of \eqref{eq:mb}-\eqref{eq:energy_balance} are not sufficient to fully characterize material behavior.
Suitable constitutive equations are needed to describe the physical phenomena in silica aerogel undergoing thermal and mechanical processes. Furthermore, the entropy production inequality (the second law of thermodynamics) imposes constraints on constitutive relations. Here, we postulate the necessary constitutive equations and refer the readers to \cite{frey1996, bowen1980, deboer2012, rajagopal1995} for the complete thermodynamical derivation of the mixture theory. 
Assuming a linear isotropic elastic \blue{for the solid constituent} results in effective solid stress as
\begin{equation}\label{eq:effstr}
\mathbf{T}'_s = 2\mu \mathbf{E}_s + \lambda \text{tr}(\mathbf{E}_s) \mathbf{I}, 
\end{equation}
where $\mathbf{E}_s = \frac{1}{2} (\nabla \mathbf{u}_s + \nabla \mathbf{u}_s^T)$ is the solid strain and computed from the the solid deformation vector $\mathbf{u}_s$ and $\lambda$ and $\mu$ are the Lam\'{e} constants that can be converted to the 
Young's modulus $E$ and Poisson's ratio $\nu$.
\blue{
We note that the linear elasticity assumption is a starting point in characterizing the mechanical behavior of novel silica aerogel.
More experimental studies are needed to understand the underlying mechanisms of deformation in 
nano-porous and thin solid walls in these porous materials to guide developing more accurate micromechanical-based models.
}
Neglecting the dilatational and temperature contributions, momentum interaction is assumed to be related to the relative movement of phases \cite{prevost1985wave, deboer2012}, 
\begin{equation}\label{eq:momentum_supply}
\mathbf{m}_s = - \mathbf{m}_f = \gamma (\mathbf{v}_s - \mathbf{v}_f),
\end{equation}
where $\gamma$ is the drag coefficient.
Since the silica aerogel phases are non-reacting, we neglect the species mass source term, $S_s = S_f = 0$. Also, a constitutive relation for the fluid density is taken into account such that,
\begin{equation}\label{eq:rhof_const}
\phi_f \frac{\partial \rho_f}{\partial p} = c_0 \rho_f,
\end{equation}
where the parameter $c_0$ is known as the \textit{constrained specific storage coefficient}, related to the bulk modulus and volume fractions of phases \cite{phillips2005finite, merxhani2016}.
Additionally, the solid and fluid partial heat fluxes are assumed to follow the Fourier law,
\begin{eqnarray}
\mathbf{q}_s &=& - \phi_s \kappa_s \nabla \theta_s,\\
\mathbf{q}_f &=& - \phi_f \kappa_f \nabla \theta_f,
\end{eqnarray}
where $\kappa_s$ and $\kappa_f$ are the solid and fluid thermal conductivities, respectively.
In porous silica aerogels, local thermal equilibrium may not be achieved due to small pore size and non-isothermal flows, leading to a significant difference between solid and fluid temperatures, see, e.g., \cite{he2015, wang2019simulation}. 
Within the mixture theory, the heat exchanges between the constituents are addressed by the internal energy supplies ${e}_\alpha$. 
For the binary mixture, we consider the below constitutive relations \cite{frey1996},
\begin{equation}
{e}_s = - {e}_f = - h (\theta_s - \theta_f).
\end{equation}
where $h>0$, in general, depends on the thermal properties and velocity fields to account for convective heat transfer.


After defining the required constitutive relations, we derive the governing equations of the silica aerogel thermo-mechanical model.
Taking into account (\ref{eq:fluid_stress}) and the momentum supplied to the fluid phase (\ref{eq:momentum_supply}), the linear momentum equation 
for fluid reduces, after neglecting the body force and inertia, to Darcy's law given by
\begin{equation}\label{eq:darcy}
\mathbf{g} = \phi_f (\mathbf{v}_f - \mathbf{v}_s) = - k \nabla p,
\end{equation}
where $\mathbf{g}$ is the fluid flux and $k = {\phi_f^2}/{\gamma}$ denotes the permeability.
Additionally, the linear momentum equation for the aerogel's solid \blue{constituent} can be re-written using 
the elasticity assumption for the effective solid stress in (\ref{eq:effstr}) as
\begin{equation}\label{eq:solid_moment}
\nabla \cdot \mathbf{T}'_s - \phi_s \nabla p + \gamma (\mathbf{v}_s - \mathbf{v}_f) = 0.
\end{equation}
Using the Darcy's law (\ref{eq:darcy}) and the fluid density constitutive relation (\ref{eq:rhof_const}), the relation (\ref{eq:gen_eq_state}) reduces to
\begin{equation}\label{eq:eqstate}
C \frac{\partial p}{\partial t} + \nabla \cdot \mathbf{v}_s - \nabla \cdot(k \nabla p) = 0,
\end{equation}
where $C$ is the fluid compressibility parameter.
Furthermore, substituting the constitutive relations and neglecting the external heat source, the governing equations of solid and fluid heat transfer are given by
\begin{eqnarray}\label{eq:heattrans}
\rho_s \phi_s c_s (\frac{\partial \theta_s}{\partial t} + \mathbf{v}_s \cdot \nabla \theta_s) &=& \mathbf{T}_s : \mathbf{D}_s + \nabla \cdot ( \phi_s \kappa_s \nabla \theta_s) - h (\theta_s - \theta_f),\\
\rho_f \phi_f c_f (\frac{\partial \theta_f}{\partial t} + \mathbf{v}_f \cdot \nabla \theta_f) &=& -\phi_f p (\text{tr} \mathbf{D}_f)+ \nabla \cdot ( \phi_f \kappa_f \nabla \theta_f) + h (\theta_s - \theta_f). \nonumber
\end{eqnarray}

In summary, the governing equations of the thermo-mechanical mixture theory consist of 
(\ref{eq:solid_moment}), (\ref{eq:eqstate}), and (\ref{eq:heattrans}).
Solving the system of partial differential equations for the fluid pressure, solid displacement, fluid temperature, and solid temperature
characterize the coupled elastic deformation and heat transfer of silica aerogel materials.

\section{Uncertainty Analysis Methods}\label{sec:uq}

Predictive computational models of physical systems with quantified uncertainty consist of two processes.
The first is the \textit{statistical forward process}, in which uncertainty in the parameters is propagated to the quantity of interests (QoIs), \blue{i.e., the targets of prediction}. The second is the \textit{statistical inverse problem}, by which the probability densities of the model inputs are learned from the observation data. 
This section is concerned with the methods employed for the treatment of uncertainty in the
\blue{thermomechanical multiphase model}
of silica aerogel materials described in section \ref{sec:mixture}. 
The UQ techniques discussed here include 
global sensitivity analysis,
Bayesian statistical inference,
and forward uncertainty propagation.

\subsection{Variance-based global sensitivity analyses}\label{sec:sensitivity}

The global sensitivity analyses of a model enable assessing the effect of model parameter uncertainty on the model outputs \cite{Saltelli2010,saltelli2008,saltelli2009,SaltelliSobol1995,Sobol1990}. 
We leverage a variance-based global sensitivity analysis (VSA) method (i.e., Sobol method) \cite{Sobol1990,Sobol1993,Sobol2007} to characterize the relative confidence in the silica aerogel model’s predictive potential. The VSA quantifies how the conditional variance caused by a parameter describes the variance in the model output. This method has many convenient features for complex physical models, including independence in evaluating the sensitivity and the ability to handle numerous model parameters all at once. We briefly describe the VSA method in this section.

Let \(Q\) be a univariate model output or the QoI and $K$ the number of uncertain parameters, \(\boldsymbol{\theta} = \{\theta_k\}_{k=1}^K\), in the model. 
The variance of the $Q$ can be decomposed such that,
\begin{equation}
 \mathbb{V}(Q) = \mathbb{V}_{\boldsymbol{\theta}_{\sim k}} (\mathbb{E}_{\theta_k}(Q|\boldsymbol{\theta}_{\sim k})) + 
 \mathbb{E}_{\boldsymbol{\theta}_{\sim k}} (\mathbb{V}_{\theta_k}(Q|\boldsymbol{\theta}_{\sim k})),
 \label{eq:modelVariance}
\end{equation}
where, $\boldsymbol{\theta}_{\sim k}$ is a vector including all input uncertain parameters except \(\theta_k\).
In this relation, 
\(\mathbb{V}_{\boldsymbol{\theta}_{\sim k}} (\mathbb{E}_{\theta_k}(Q|\boldsymbol{\theta}_{\sim k}))\) is the reduction in variance of \(Q\) when all parameters except \(\theta_k\) are kept constant and \(\mathbb{E}_{\boldsymbol{\theta}_{\sim k}} (\mathbb{V}_{\theta_k}(Q|\boldsymbol{\theta}_{\sim k}))\) is the residual variance of \(Q\) when \(\theta_k\) is kept constant.
The total effect sensitivity index \cite{HommaSaltelli1996,SaltelliTarantola2002} is then defined as 
\begin{equation}
 \mathcal{S}_k = \frac{\mathbb{E}_{\boldsymbol{\theta}_{\sim k}} (\mathbb{V}_{\theta_k}(Q|\boldsymbol{\theta}_{\sim k}))}{\mathbb{V}(Q)}=1-\frac{\mathbb{V}_{\boldsymbol{\theta}_{\sim k}} (\mathbb{E}_{\theta_k}(Q|\boldsymbol{\theta}_{\sim k}))}{\mathbb{V}(Q)}.
 \label{eq:total effect index}
\end{equation}
The total effect index estimates the effect of the input \(\theta_k\) to the variation of the model output. Thus, considering a constant value for a parameter with a small total effect index in the range of uncertainty will not significantly impact the model output.
To this end, parameter sensitivity can determine the most influential parameters of a model and guide reducing the parameter space for inference.

\subsubsection{Numerical estimator of total effect sensitivity index}
Saltelli \cite{HommaSaltelli1996,Saltelli2002,Saltelli2010,saltelli2008} suggested an efficient Monte-Carlo estimator to evaluate the total sensitivity index $\mathcal{S}$. The estimator consists of constructing two \(N \times K\) matrices, \(\boldsymbol{A}\) and \(\boldsymbol{B}\), by drawing \(N\) random samples from each uncertain parameter probability distribution. The matrices \({\boldsymbol{A}}_{\boldsymbol{B}}^{(k)}\), \(k=1,2,...,K\) are then built from all columns of \(\boldsymbol{A}\), excluding the \(k\)th, that comes from $\boldsymbol{B}$. 
The vectors \({\boldsymbol{y}}_{\boldsymbol{A}}\) and \({\boldsymbol{y}}_{\boldsymbol{AB}}^{(k)}\) include the model outputs in each row of \(\boldsymbol{A}\) and \({\boldsymbol{A}}_{\boldsymbol{B}}^{(k)}\). 
The total-effect index for \(\theta_k\) is approximated by \cite{Saltelli2010}:
\begin{equation}\label{eq:mc_sensitivity}
 \mathcal{S}_k \approx \frac{1}{2N}\sum_{j=1}^N\left (({\boldsymbol{y}}_{\boldsymbol{A}})_j-({\boldsymbol{y}}_{\boldsymbol{AB}}^{(k)})_j\right)^2.
\end{equation}
The Monte-Carlo estimator of \(\mathcal{S}_k\) in (\ref{eq:mc_sensitivity}) reduces the cost of evaluating the multi-dimensional
integrals from \(N^2\) to \(N(K+2)\) model evaluations \cite{Saltelli2010}.

\subsection{Bayesian learning from uncertain data}\label{sec:bayes}
An essential process in predictive computational modeling is using a learning algorithm to train the model by use of observational data. Predictive modeling of physical systems poses an additional challenge due to the presence of uncertainty arising from incompleteness and noise in the data and modeling error, which translates to uncertainty in model prediction.
We employ a Bayesian learning approach to the statistical inference problem, as it offers a self-consistent framework for predictive physics-based modeling that enables characterizing uncertainty in data, model parameters, and the model itself \cite{dehghannasiri2017,honarmandi2017,honarmandi2020,kim2020,singer2013}. This section summarizes the Bayesian approach for learning the physics-based model from uncertainty data as described in \cite{faghihi2018fatigue,farrell2015jcp,odenbabuska2017,oden2015amses,prudencio2015,prudencio2014CompB}.

Let \(\boldsymbol{\theta}\) be a vector of model parameters belonging to a parameter space \(\Theta\) and \(\boldsymbol{D}\) be the training data belonging to a space of observational data \(\mathcal{D}\). In the Bayesian setting, the model parameters and training data are both random variables characterized by probability density functions (PDFs), \(\pi(\boldsymbol{\theta})\) and \(\pi(\boldsymbol{D})\). 
Training the computational model consists of learning the model parameters from data. Then, following a statistical inference method, the trained model parameters are obtained through the Bayes’ theorem \cite{jaynes2003}
\begin{equation}
 \pi_{post}(\boldsymbol{\theta}|\boldsymbol{D}) = \frac{\pi_{like}(\boldsymbol{D}|\boldsymbol{\theta})\pi_{prior}(\boldsymbol{\theta)}}{\pi_{evid}(\boldsymbol{D})},
 \label{eq:bayes}
\end{equation}
where \( \pi_{post}(\boldsymbol{\theta}|\boldsymbol{D})\) is the posterior PDF, which defines the Bayesian update of the prior information represented by \(\pi_{prior}(\boldsymbol{\theta)}\),
 \(\pi_{like}(\boldsymbol{D}|\boldsymbol{\theta})\) is the likelihood PDF, 
 and \(\pi_{evid}(\boldsymbol{D})\) is the evidence, which is the probability of observed data,
\begin{equation}
 \pi_{evid}(\boldsymbol{D}) = \int \pi_{like}(\boldsymbol{D|\boldsymbol{\theta}}).\pi_{prior}(\boldsymbol{\theta})d\boldsymbol{\theta}.
\end{equation}
The prior PDF in Bayesian calibration (\ref{eq:bayes}) represents our initial knowledge of the model parameters. Based on available features of the model parameter (i.e., bounds, mean, and variance), the \textit{maximum entropy principle} can be used to create the prior \cite{jaynes1982,kaipio2006}. If only parameter bounds are available, i.e., total ignorance, uniform distribution should be used as the parameter prior \cite{jaynes2003}. 

%
The statistical discrepancy between the model output \(\boldsymbol{d}(\boldsymbol{\theta})\) and the observational data \(\boldsymbol{D}\), known as noise model, defines the form of the likelihood, \(\pi_{like}(\boldsymbol{D|\boldsymbol{\theta}})\) in (\ref{eq:bayes}). 
Let \(p_{\epsilon}\) be a probability distribution for the total error \(\boldsymbol{\epsilon} = \boldsymbol{\eta} + \boldsymbol{\xi}(\boldsymbol{\theta}) = \boldsymbol{D}-\boldsymbol{d}(\boldsymbol{\theta})\) due to modeling errors, \(\boldsymbol{\xi}(\boldsymbol{\theta})\), as well as data noise \(\boldsymbol{\eta}\). 
\blue{Here, in the spirit of Kennedy and O'Hagan \cite{kennedy2001bayesian}, an additive noise model is assumed to simultaneously account for both data noise and modeling errors. In particular, 
the modeling errors stem from the inadequacy of the theoretical model developed in Section \ref{sec:mixture} in capturing complex physical processes, as well as the error incurred by numerical solution approximations such as discretization errors and two-dimensional (2D) assumption for the inherently three-dimensional (3D) physical phenomena.}
We consider the total error to be a Gaussian random variable with a mean of zero, \(\boldsymbol{\epsilon} \sim \mathcal{N} (0,\boldsymbol{\Gamma}_{noise}^{-1})\), where \(\boldsymbol{\Gamma}_{noise}\) is the noise covariance matrix. 
Thus, the probability density function of the likelihood is,
\begin{equation}
 \pi_{like}(\boldsymbol{D|\boldsymbol{\theta}}) = p_{\epsilon}(\boldsymbol{D}-\boldsymbol{d}(\boldsymbol{\theta})).
\end{equation}
Assume that $N_D$ independent and identically distributed (i.i.d.) data $\boldsymbol{D}^{(j)}$ 
have been sampled from the probability distribution \( \boldsymbol{D}^{(j)} \sim p(\boldsymbol{D})\),
and \(\boldsymbol{d}_i(\boldsymbol{\theta})\) is the corresponding model output. 
The likelihood function is then represented by,
\begin{eqnarray}\label{eq:likelihood}
 \ln\left(\pi_{like}(\boldsymbol{D}|\boldsymbol{\theta})\right) 
 & = & 
 \frac{1}{2}\sum_{j=1}^{N_D} \left(\boldsymbol{d}(\boldsymbol{\theta})-\boldsymbol{D}^{(j)} \right)^T 
 \boldsymbol{\Gamma}_{noise}^{-1}
 \left(\boldsymbol{d}(\boldsymbol{\theta})-\boldsymbol{D}^{(j)} \right) \nonumber \\
 & + & \frac{N_tN_D}{2} \ln(2\pi) + \frac{1}{2} \ln | \boldsymbol{\Gamma}_{noise} |, 
\end{eqnarray}
where $N_t$ is the number of data points
and $\| \boldsymbol{d}(\boldsymbol{\theta})-\boldsymbol{D}^{(j)} \|^2$ is known as data mistfit.
The noise covariance matrix $\boldsymbol{\Gamma}_{noise}$ describes both data noise $\boldsymbol{\eta}$ and error in modeling assumptions $\boldsymbol{\xi}(\boldsymbol{\theta})$.
The ultimate success of Bayesian inference in predictive modeling depends upon how well the noise covariance is characterized.
Overestimated noise levels may lead to discarding useful information contained in the observational data. On the other hand, 
underestimating the noise will result in overfitting the model to measurement and model errors, leading to overconfident and highly biased estimates.
To avoid biasing the Bayesian inference, we parametrize the noise model and infer the resultant unknown hyper-parameter.
We assume that modeling errors are proportional to measurements errors. That is,
\begin{equation}\label{eq:noise}
 \boldsymbol{\Gamma}_{noise} = m (\sigma_i)^2 \mathbf{I}, \quad i=1,\cdots,N_t,
\end{equation}
where $(\sigma_i)^2$ is the variance of each observational data, and the unknown multiplier $m$ characterizes the modeling error.
We view the noise multiplier $m$ as a hyper-parameter to be calculated in the Bayesian inference process.
A careful choice of a mathematically compatible prior for the noise multiplier
$m$ is crucial to inference success. We thus chose
the inverse gamma function with the mean of 1.0 and the standard deviation of 0.1 \cite{dakota2020}.

\subsection{Computational solution of Bayesian inference}\label{sec:mcmc}

The Bayesian learning of physics-based models requires computing the posterior PDFs of the parameters $\pi_{post}(\boldsymbol{\theta}|\boldsymbol{D})$ as the solution of the statistical inference problem according to (\ref{eq:bayes}). 
Markov Chain Monte Carlo (MCMC) sampling methods are typically used to draw samples from the parameter posteriors given the prior PDFs, and the form of the likelihood function \cite{gelman1992,roberts2004}.  
The MCMC solution of the Bayesian inference for the forward models governed by nonlinear systems of partial differential equations (such as the mixture thermo-mechanical model described in section \ref{sec:mixture}) is computationally expensive due to the large number of model evaluations required to explore the posterior distribution.
Common MCMC method is the Metropolis-Hastings (MH) algorithm \cite{metropolis1953,hastings1970}, in which 
an initial parameter value $\boldsymbol{\theta}^{(0)}$ is specified and at the \(l\)-th iteration, a candidate \(\boldsymbol{\theta}^{*}\) is chosen from a proposal distribution \(q(\boldsymbol{\theta}^{*}|\boldsymbol{\theta}^{(l)})\).
Often, \(q(\boldsymbol{\theta}^{*}|\boldsymbol{\theta}^{(l)})\) follows a Gaussian distribution with a mean $\boldsymbol{\theta}^{(l)}$ and a fixed covariance.
At the $(l + 1)$ step in the chain, the candidate sample $ \boldsymbol{\theta}^*$ is accepted, $\boldsymbol{\theta}^{(l+1)} = \boldsymbol{\theta}^*$,
 with the probability of $min\{1,\alpha(\boldsymbol{\theta}^*,\boldsymbol{\theta}^{(l)})\}$, otherwise it is rejected, $\boldsymbol{\theta}^{(l+1)} = \boldsymbol{\theta}^l$.
The acceptance ratio \(\alpha\) is computed as (see \cite{kaipio2006} for more details)
\begin{equation}
 \alpha(\boldsymbol{\theta}^*,\boldsymbol{\theta}^{(l)})=\frac{\pi_{post}(\boldsymbol{\theta}^*)q(\boldsymbol{\theta}^{(l)}|\boldsymbol{\theta}^*)}{\pi_{post}(\boldsymbol{\theta}^{(l)})q(\boldsymbol{\theta}^{*}|\boldsymbol{\theta}^{(l)})}.
\end{equation}

There are several drawbacks associated with the MH sampler.
Many samples will be rejected if the proposal variance is chosen too high leading to poor sampling, and the proposal distribution might not adequately represent the local shape of the target posterior distribution. 
The small proposal variance will result in higher acceptance, while the higher likelihood
regions in the posterior are explored and miss the tails. 
In this project, we make use of
Delayed Rejection Adaptive Metropolis (DRAM) algorithm as implemented in \cite{dakota2020} to solve the Bayesian inference.
DRAM improves the MH sampler by testing a few backup samples with smaller proposal variance before rejecting the candidate sample (i.e., Delay Rejection scheme).
Additionally, the proposal covariance is adapted to match the posterior covariance in DRAM sampler (i.e., Adaptive Metropolis scheme).

\subsection{Solution of the statistical forward problem}\label{sec:stat_forward}

Once the model parameters are learned from observational data through Bayesian inference, one can assess the reliability of the computational model in predicting the QoIs. The parameter posteriors characterize the uncertainty (lack of knowledge or error) in the model and data. To assess the probability distribution of the QoIs, such uncertainties in the parameters must be propagated through the model solution. 
The process requires the solution of the statistical forward problem, using Monte Carlo (MC) method. Starting from an initial point $\boldsymbol{\theta}^{(0)}$, the sequence a random walk generates a random number. A good choice of the initial point is the Maximum A Posteriori estimate that is a point estimate defined as
\begin{equation}
 \boldsymbol{\theta}^{MAP} = \operatorname*{argmin}_\theta \pi_{post}(\boldsymbol{\theta|\boldsymbol{D}}),
 \label{eq:MAP}
\end{equation}
and can be estimated using samples of posterior distributions obtained using an MCMC algorithm.
The MC sampler continues by generating the next random number by choosing a trial $\boldsymbol{\theta}^{*}$ drawn from a uniform distribution with the mean, centered at the value of the current state and the range of fixed maximum step size. We accept the trial if it is towards a higher probability region and conditionally reject it if the trial is toward a lower probability region. Like the MCMC algorithm, the step size must be cautiously chosen since a small step size leads to very slow convergence as most of the trial steps will be accepted. On the other hand, if the step size is too large, the random walker might not visit the posterior distribution’s critical peaks.

We conclude this section by specifying the metric we use in Section \ref{sec:results} to compare the data and posterior model prediction obtained from the MC sampler.
Let \(\Pi^D(Q))\) and \(\Pi^D(d))\) be the QoI's cumulative distribution functions derived from the model, \(Q^d\), and the data, \(Q^D\), respectively. The measure to access how well the model is representing the data is,
\begin{equation}\label{eq:error}
 \epsilon = \frac{\int_{-\infty}^{\infty}\left|\Pi^D(\xi)-\Pi^d(\xi) \right|d\xi}{\mathbb{E}(Q^D)}
\end{equation}
where the denominator is the mean of the \(Q^D\), obtained from sampler estimation.

\section{Results}\label{sec:results}

\subsection{Experimental data of silica aerogel}\label{sec:data}

\begin{figure}[h!]
 \centering
 \subfloat[]{\includegraphics[width=0.45\textwidth]{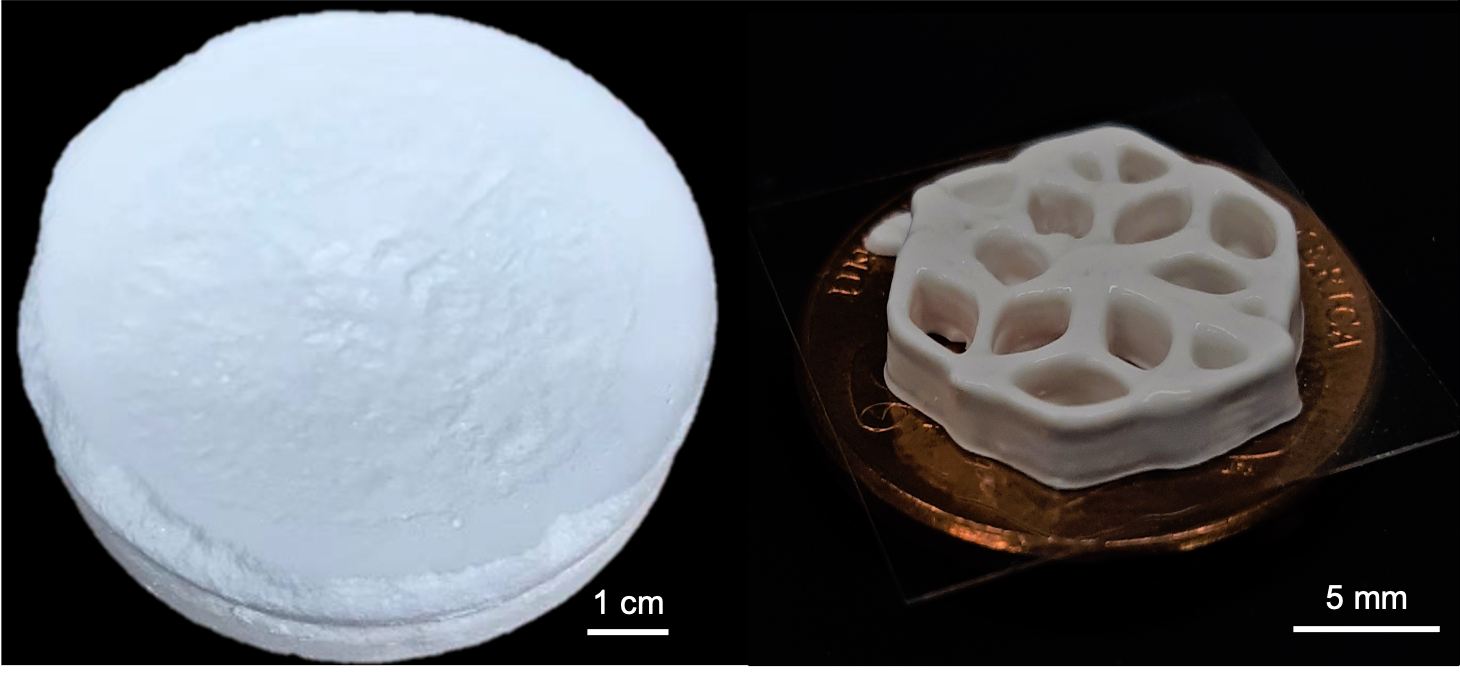}
 }
 \hfill
 \subfloat[]{\includegraphics[width=0.53\textwidth]{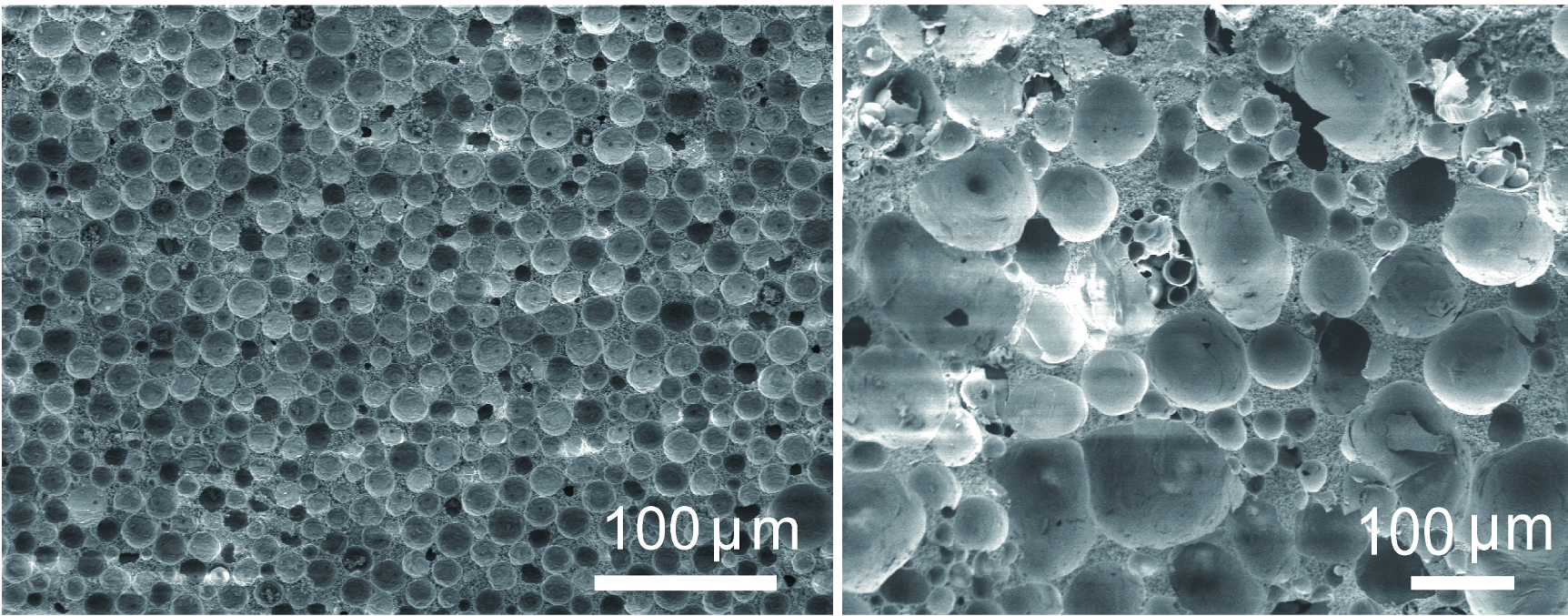}
 }
 \caption{
 \blue{
 Silica aerogel fabricated by in-situ pore formation and ambient pressure and temperature drying:
 (a) The optical image of a one-pot synthesized sample and an additive manufactured part using DIW,
 (b) Scanning electron microscopy images of the aerogel with different porosities \cite{yang2019hierarchical, chi2021}.
 }}
 \label{fig:aerogel}
\end{figure}

The standard aerogel fabrication is the supercritical drying process, where the highly porous materials are produced by removing the solvent without collapsing the solid structure.
Supercritical drying suffers from substantial energy consumption, long processing time, high-cost equipment, and small-dimension production \cite{wei2007, van1995}.
We have recently developed a new synthesis method for silica aerogel that relies on an in-situ bubble-supported pore formation.  
In this approach, the silica aerogel is prepared using a mixture of tetraethyl orthosilicate (TEOS), cetrimonium bromide (CTAB), urea (foaming agent), acetic acid, and distilled water. The porous structure is formed by the thermal decomposition of urea to \(CO_2\), and \(NH_3\) bubbles. Combined with ambient pressure and temperature drying, this synthesis method \blue{reduces the cost of aerogel production by 60\%} \cite{yang2019hierarchical}.
We have also demonstrated the additive manufacturing of silica aerogel into the customized parts using a direct ink writing (DIW) method.
The printable aerogel ink with embedded gaseous bubbles is facilitated by modifying the rheological properties of the silica aerogel. In particular, the cellulose-based viscosity modifier is used to achieve the highly viscous and shear-thinning non-Newtonian aerogel ink enabling the 3D printing (see Figure \ref{fig:aerogel}) \cite{chi2021}.

\begin{figure}[h!]
 \centering
 \vspace{-0.15in}
 \subfloat[]{\includegraphics[width=0.48\textwidth]{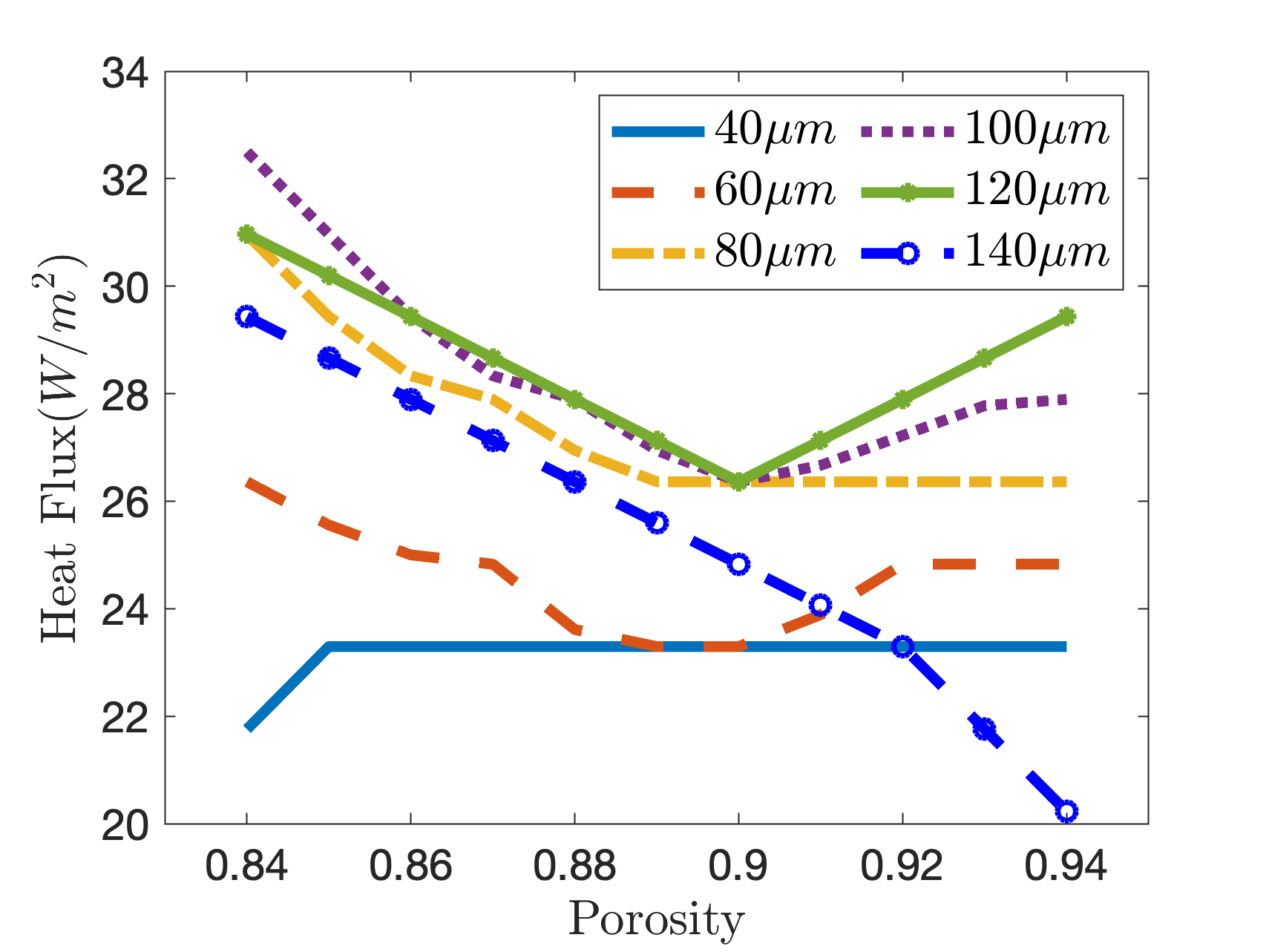}
 }
 \hfill
 \subfloat[]{\includegraphics[width=0.48\textwidth]{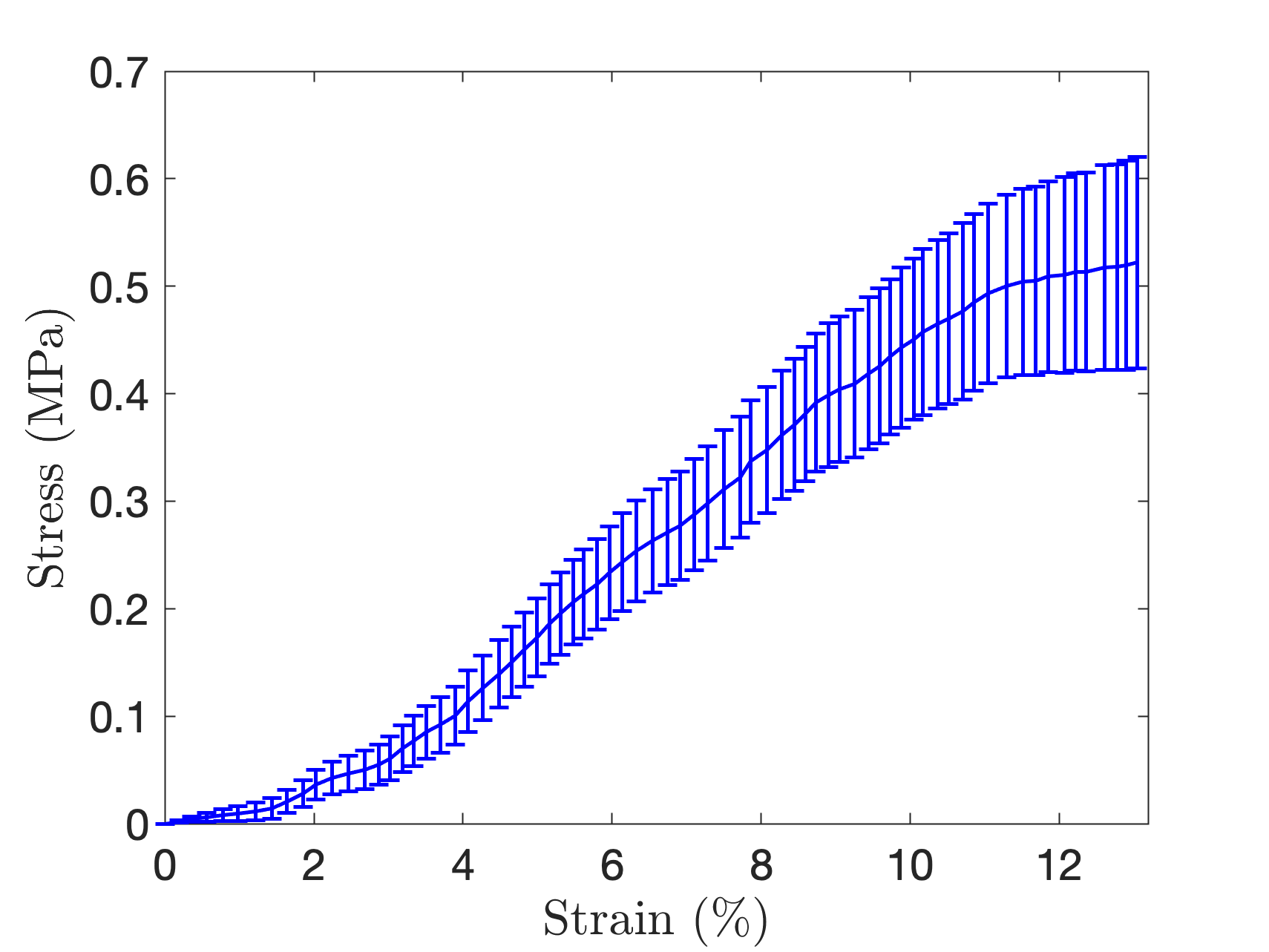}
 }
 \vspace{-0.1in}
 \caption{
 Experimental data of thermal and mechanical properties of silica aerogel:
 (a) Heat flux measurements for different aerogel porosities \blue{(0.84 - 0.94) and pore sizes (40 and 140 \(\mu m\))},
 (b) Stress-strain measurements of 3D printed aerogel under uniaxial compression \cite{yang2019hierarchical, chi2021}.
 }
 \label{fig:exp_data}
\end{figure}

Figure \ref{fig:exp_data} shows the experimental characterization of thermal and mechanical properties of silica aerogel \cite{yang2019hierarchical, chi2021}, employed in this work for the Bayesian calibration of the multiphase model described in Section \ref{sec:mixture}.
Following the ASTM C518 standard, the thermal measurements of the silica aerogel in Figure \ref{fig:exp_data}(a) consists of subjecting 
aerogel samples with the dimension of 1$\times$1$\times$0.6 cm to a temperature $\bar{\theta}_{hot} = 29.5^\circ C$ at the top boundary and the heat flux at the bottom boundary is measured using a heat flux sensor (see Figure \ref{fig:BC})
The heat flux measurements in Figure \ref{fig:exp_data}(a) corresponds to the silica aerogel samples with different pore sizes.
A 3D printed sample with a dimension of 10$\times$10$\times$10 mm was utilized in a uniaxial compression test to characterize the stress-strain relationship depicted in Figure \ref{fig:exp_data}(b).
Error bars are added to represent the data uncertainty in material characterization (experimental noise).
\blue{
Note that the heat flux data in Figure \ref{fig:exp_data}(a) were obtained from a single sample for each porosity level. Thus, no information regarding measurement variability is available for the inversion. Nevertheless, in Section \ref{sec:calibration}, we use these measurements to characterize microstructural uncertainty due to pore size variabilities
that cannot be captured by the multiphase model.}

\subsection{Computational models of silica aerogel }\label{sec:fe}

\begin{figure}[h!]
 \centering
 \includegraphics[width= 4in]{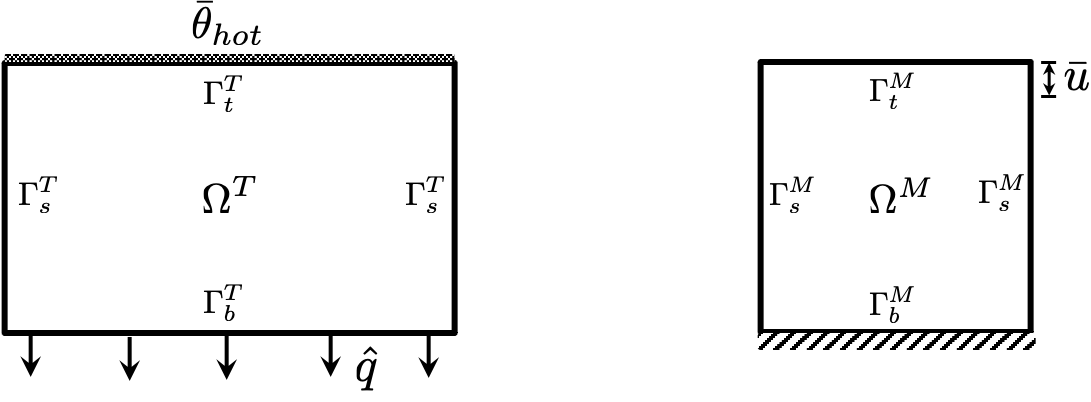}
 \\(a) \hspace{2in} (b)\\
 \caption{Domains and boundary conditions for modeling the thermal and mechanical experimental measurements of silica aerogel.
 (a) Thermal model in which $\bar{\theta}_{hot}$ is imposed at the top surface of the specimen and the heat flux is measured in the bottom surface.
 (b) Mechanical model corresponding to the uniaxial compression test with the permeable boundary for the fluid phase.}
	\label{fig:BC}
\end{figure}

We simulate the thermal and mechanical behavior of silica aerogel using the two-dimensional (2D) multiphase models of silica aerogel derived in Section \ref{sec:mixture}.
The domains and boundary conditions corresponding to the heat transfer thermal and mechanical experimental measurements are shown in Figure \ref{fig:BC}.
The governing equations and the corresponding boundary conditions of the two-phase heat transfer model are
\begin{equation}\label{eq:thermal_bcs}
\left.\begin{aligned}
\rho_s \phi_s c_s \frac{\partial \theta_s}{\partial t} = \nabla \cdot ( \phi_s \kappa_s \nabla \theta_s) - h (\theta_s - \theta_f) \hspace{0.5cm} \text{in} \hspace{0.5cm} \Omega^T\\ 
\rho_f \phi_f c_f \frac{\partial \theta_f}{\partial t} = \nabla \cdot ( \phi_f \kappa_f \nabla \theta_f) + h (\theta_s - \theta_f)
\hspace{0.5cm} \text{in} \hspace{0.5cm} \Omega^T\\ 
 \theta_s = \theta_f = \bar{\theta}_{hot} \hspace{0.5cm} \text{on} \hspace{0.5cm} \Gamma^T_t\\ 
 -\phi_s \kappa_s\nabla \theta_s = h_{air}(\theta_s-\bar{\theta}_{cold}) \hspace{0.5cm} \text{on} \hspace{0.5cm} \Gamma^T_b\\
 -\phi_f \kappa_f \nabla \theta_f =h_{air}(\theta_f-\bar{\theta}_{cold}) \hspace{0.5cm} \text{on} \hspace{0.5cm} \Gamma^T_b\\
 -\phi_s \kappa_s\nabla \theta_s = - \phi_f \kappa_f\nabla \theta_f = 0 \hspace{0.5cm} \text{on} \hspace{0.5cm} \Gamma^T_s\\
\end{aligned}\right\}.
\end{equation}
Compared to \eqref{eq:heattrans}, the mechanically coupled terms are ignored in \eqref{eq:thermal_bcs}
since there is no mechanical stress in the experimental thermal measurements.
Additionally, it is assumed that the velocities of the solid and fluid phases are negligible 
($\mathbf{v}_f \approx 0$ and $\mathbf{v}_s \approx 0$), 
since pores form small unconnected structures (see microstructure images in \cite{yang2019hierarchical}).
The boundary conditions in \eqref{eq:thermal_bcs}, replicates the thermal measurements in which the $\bar{\theta}_{hot}$ is imposed to the top boundary, 
convective heat transfer with coefficient $h_{air}$ and temperature $\bar{\theta}_{cold}$ is considered at the bottom boundary,
and zero heat flux is applied to the side boundaries.
The observable heat flux $\hat{q}$ of the mixture, corresponding to the flux measurements in the experiments, is computed using the third relation in \eqref{eq:mix_sum}.

The governing equations and the corresponding boundary conditions of the two-phase deformation model of aerogel are
\begin{equation}\label{eq:mech_bcs}
\left.\begin{aligned}
C \frac{\partial p}{\partial t} + \frac{\partial }{\partial t}(\nabla \cdot \mathbf{u}_s) + \nabla \cdot \mathbf{g} = 0
\hspace{0.5cm} in \hspace{0.5cm} \Omega^M\\ 
\mathbf{g} = -k \nabla p \hspace{0.5cm} in \hspace{0.5cm} \Omega^M\\
\nabla \cdot \mathbf{T}_s (\mathbf{u}_s) = 0 \hspace{0.5cm} in \hspace{0.5cm} \Omega^M\\ 
 \mathbf{u}_s = \bar{\mathbf{u}}_s \hspace{0.5cm} on \hspace{0.5cm} \Gamma^M_t\\ 
 \mathbf{u}_s = \mathbf{0} \hspace{0.5cm} on \hspace{0.5cm} \Gamma^M_b\\
 p = 0 \hspace{0.5cm} on \hspace{0.5cm} \Gamma^M\\
 \mathbf{g} = \mathbf{0} \hspace{0.5cm} on \hspace{0.5cm} \Gamma^M\\  
\end{aligned}\right\},
\end{equation}
where a three-field formulation, similar to the poroelastic model, e.g., \cite{phillips2009, haagenson2020}, is used with state variables given by the fluid pressure $p$, fluid flux $\mathbf{g}$, and solid displacement $\mathbf{u}_s$.
For modeling the compression test, quasi-static conditions are assumed, and deformation $\bar{{u}}$ at a constant strain rate is applied incrementally to the top boundary.
The fluid pressure is assumed to be zero in all boundaries, representing the fully permeable surfaces of the silica aerogel samples.

\subsubsection{Finite element solutions}

We make use of continuous Galerkin finite element methods to solve the coupled system of equations of the heat transfer \eqref{eq:thermal_bcs} and deformation \eqref{eq:mech_bcs} models.
%
%
The heat transfer model involves the solid temperature $\theta_s$ and fluid temperature $\theta_f$ as the unknown state variables.
To define the relevant finite element spaces, we consider the Hilbert spaces
\begin{eqnarray}
\mathcal{Z} &=& \{z \in H^1(\Omega) : z = \bar{z} \;\; \text{on} \;\; \Gamma \},\\
\mathcal{Z}^0 &=& \{z \in H^1(\Omega) : z = 0 \;\; \text{on} \;\; \Gamma \}.
\end{eqnarray}
The variational formulation for the heat transfer model, is defined then as:

\noindent Find $(\theta_s, \theta_f) \in \mathcal{Z}\times\mathcal{Z}$, such that
\begin{equation}\label{eq:thermal_var}
\left.\begin{aligned}
\int_{\Omega^T} \rho_{s} \phi_s c_{s} \theta_s^{n+1} z_s \; d\mathbf{x}
+ \int_{\Omega^T} \Delta t \; \phi_s \kappa_s (\nabla \theta_s^{n+1} \cdot \nabla z_s) \; d\mathbf{x} 
- \; \int_{\Omega^T} \Delta t \; h (\theta_f^{n+1} - \theta_s^{n+1}) z_s \; d\mathbf{x}\\
+ \int_{\Gamma^T} \Delta t \; \phi_s h_{air} (\theta_s^{n+1} - \theta_{cold}) z_s \; d\mathbf{s}
 = \int_{\Omega^T} \rho_{s} \phi_s c_{s} \theta_s^{n} z_s \; d\mathbf{x}
\\ 
\int_{\Omega^T} \rho_{f} \phi_f c_{f} \theta_f^{n+1} z_f \; d\mathbf{x}
+ \int_{\Omega^T} \Delta t \; \phi_f \kappa_f (\nabla \theta_f^{n+1} \cdot \nabla z_f) \; d\mathbf{x}
+ \int_{\Omega^T} \Delta t \; h (\theta_f^{n+1} - \theta_s^{n+1}) z_f \; d\mathbf{x}\\
+ \int_{\Gamma^T} \Delta t \; \phi_f h_{air} (\theta_f^{n+1} - \theta_{cold}) z_s \; d\mathbf{s}
 = \int_{\Omega^T} \rho_{f} \phi_f c_{f} \theta_f^{n} z_f \; d\mathbf{x}\\ 
\end{aligned}\right\},
\end{equation}
for all choices of test functions $z_s, z_f \in \mathcal{Z}^0$, where $\Delta t$ is the time increments and $n$ indicates the time step.
We employ a dual-mixed finite element method to discretize the two-field formulation in \ref{eq:thermal_var},
where $\theta_s$ and $\theta_f$ are approximated by the first-order Lagrange polynomial.
The system of nonlinear equations arising in \eqref{eq:thermal_var} is solved using a Newton-type numerical algorithm.


With respect to the finite element discretization of the deformation model, we consider the following function spaces,
\begin{eqnarray}
\mathcal{W}_p = \mathcal{W}^0_p &=& \{ p \in L^2(\Omega)\}, \nonumber\\
\mathcal{W}_g = \mathcal{W}^0_g &=& \{ \mathbf{g} \in H(\text{div};\Omega) : \mathbf{g} \cdot \mathbf{n} = 0 \;\;\text{on}\;\; \Gamma\}, \nonumber\\
\mathcal{W}_u &=& \{ \mathbf{u} \in H^1(\Omega) : \mathbf{u} = \bar{\mathbf{u}} \;\;\text{on}\;\; \Gamma\},\nonumber\\
\mathcal{W}^0_u &=& \{ \mathbf{u} \in H^1(\Omega) : \mathbf{u} = \mathbf{0} \;\;\text{on}\;\; \Gamma\}.
\end{eqnarray}
where $H(\text{div};\Omega) = \{f: f \in L^2(\Omega) \; \text{and} \; \nabla \cdot f \in L^2(\Omega) \}$. This choice guarantees the well-posedness of the variational form of the three-field deformation model.
Using the backward Euler time discretization,
the variational problem of the system of equations in \ref{eq:mech_bcs}, 
is defined as:

\noindent Find $(p, \mathbf{g}, \mathbf{u}_s) \in \mathcal{W}_p\times\mathcal{W}_g\times\mathcal{W}_u$, such that
\begin{equation}\label{eq:mech_var}
\left.\begin{aligned}
\int_{\Omega^M} \left( C p^{n+1} + \nabla \cdot \mathbf{u}_s^{n+1} \right) w_p \; d\mathbf{x} + 
\int_{\Omega^M} \Delta t (\nabla \cdot \mathbf{g}^{n+1}) w_p \; d\mathbf{x} &= 
\int_{\Omega^M} \left( C p^{n} + \nabla \cdot \mathbf{u}_s^{n} \right) w_p \; d\mathbf{x} ,
\\
- \int_{\Omega^M} \frac{\Delta t}{k} \mathbf{g}^{n+1} \cdot \boldsymbol{w}_g \; d\mathbf{x} +
\int_{\Omega^M} \Delta t p^{n+1} \nabla \cdot \boldsymbol{w}_g \; d\mathbf{x} 
&= 0,
\\  
 \int_{\Omega^M} \mathbf{T}_s(\mathbf{u}_s^{n+1}) : \nabla \boldsymbol{w}_u \; d\mathbf{x}
& = 0
\end{aligned}\right\},
\end{equation}
for all choices of test functions $w_p \in \mathcal{W}^0_p$, $\boldsymbol{w}_g \in \mathcal{W}^0_g$, and $\boldsymbol{w}_u \in \mathcal{W}^0_u$. 
Following the numerical solution of the poroelastic model in \cite{ferronato2010}, we employ a mixed finite element method to discretize the three-field formulation in \eqref{eq:mech_var}. The fluid pressure $p$ is approximated by piecewise constant functions, the flux $\mathbf{g}$ is approximated in the lowest-order Raviart-Thomas space, and the solid displacement $\mathbf{u}_s$ is approximated by the second-order Lagrange polynomial. Such a discretization allows for conserving mass discretely and alleviating the instability and pressure oscillations that can form in the solution of this class of coupled partial differential equations, e.g., \cite{phillips2009, haga2012}.


%
\begin{figure}[h!]
 \centering
 \vspace{-0.1in}
 \subfloat[]{\includegraphics[width= 0.5\textwidth]{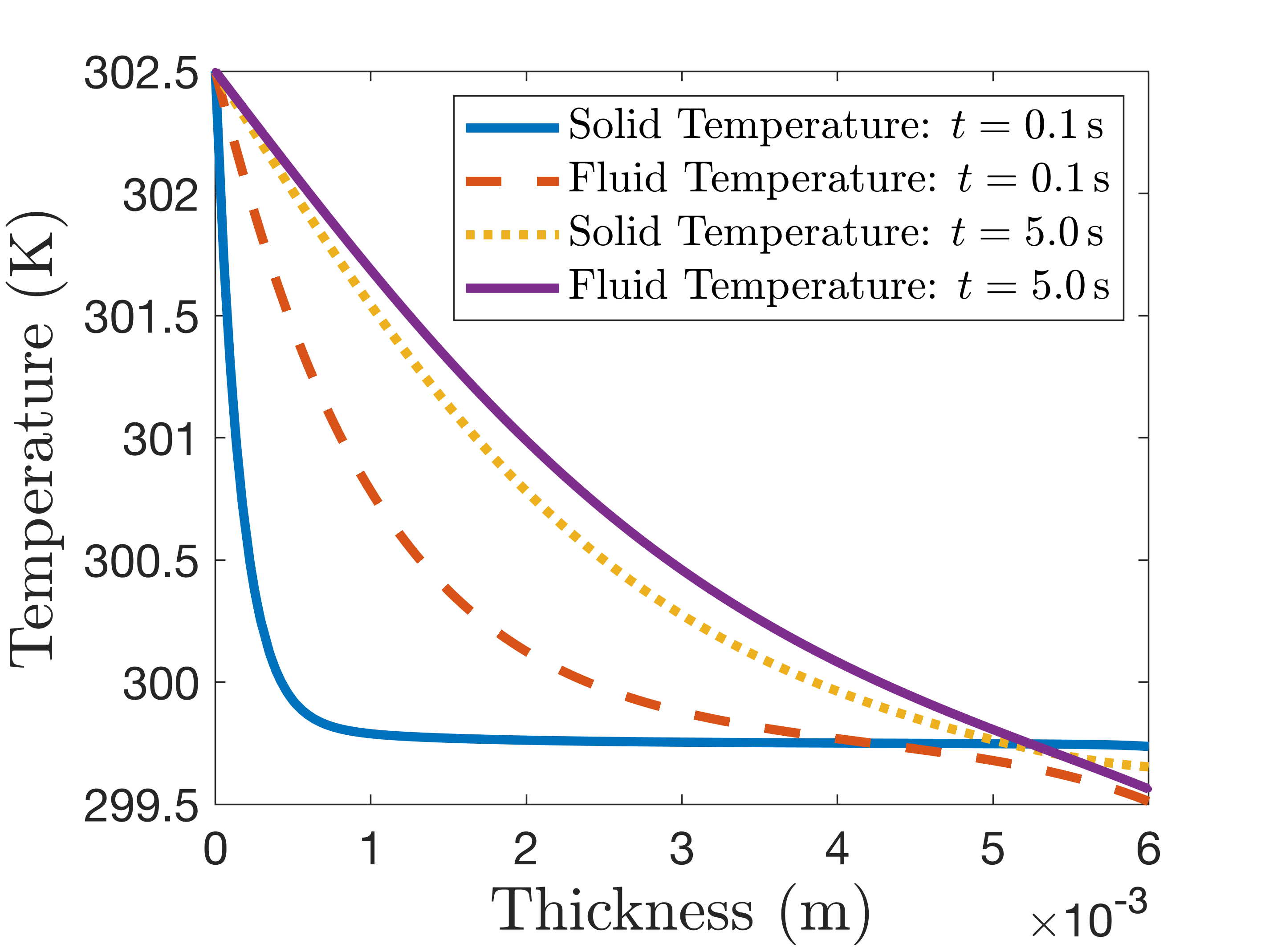}
 }
 \\\vspace{-0.1in}
 \subfloat[]{\includegraphics[trim=2.0in 3.5in 2.0in 0.0in, clip, width= 0.35\textwidth]{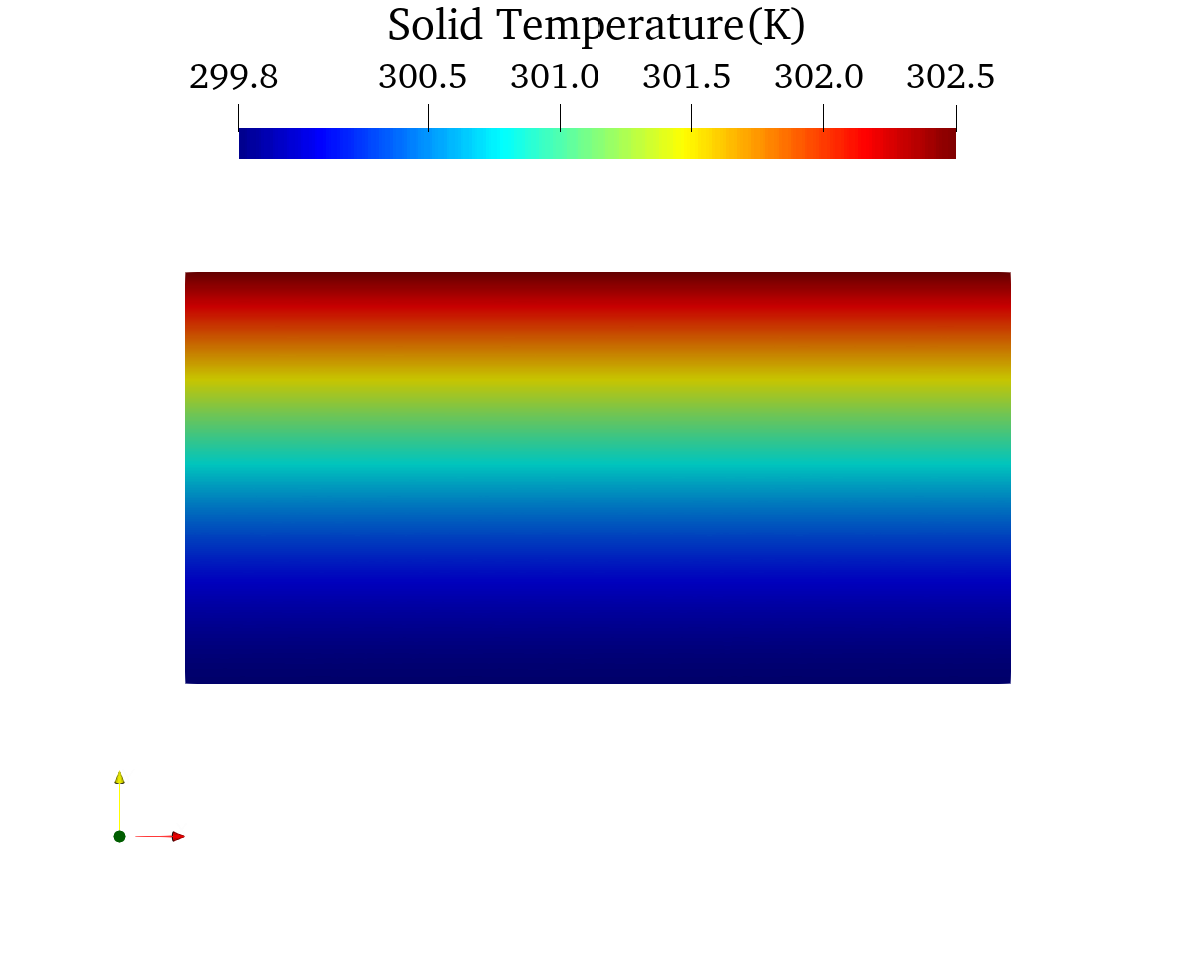}
 }
 \hspace{0.1in}
 \subfloat[]{\includegraphics[trim=2.0in 3.5in 2.0in 0.0in, clip, width= 0.35\textwidth]{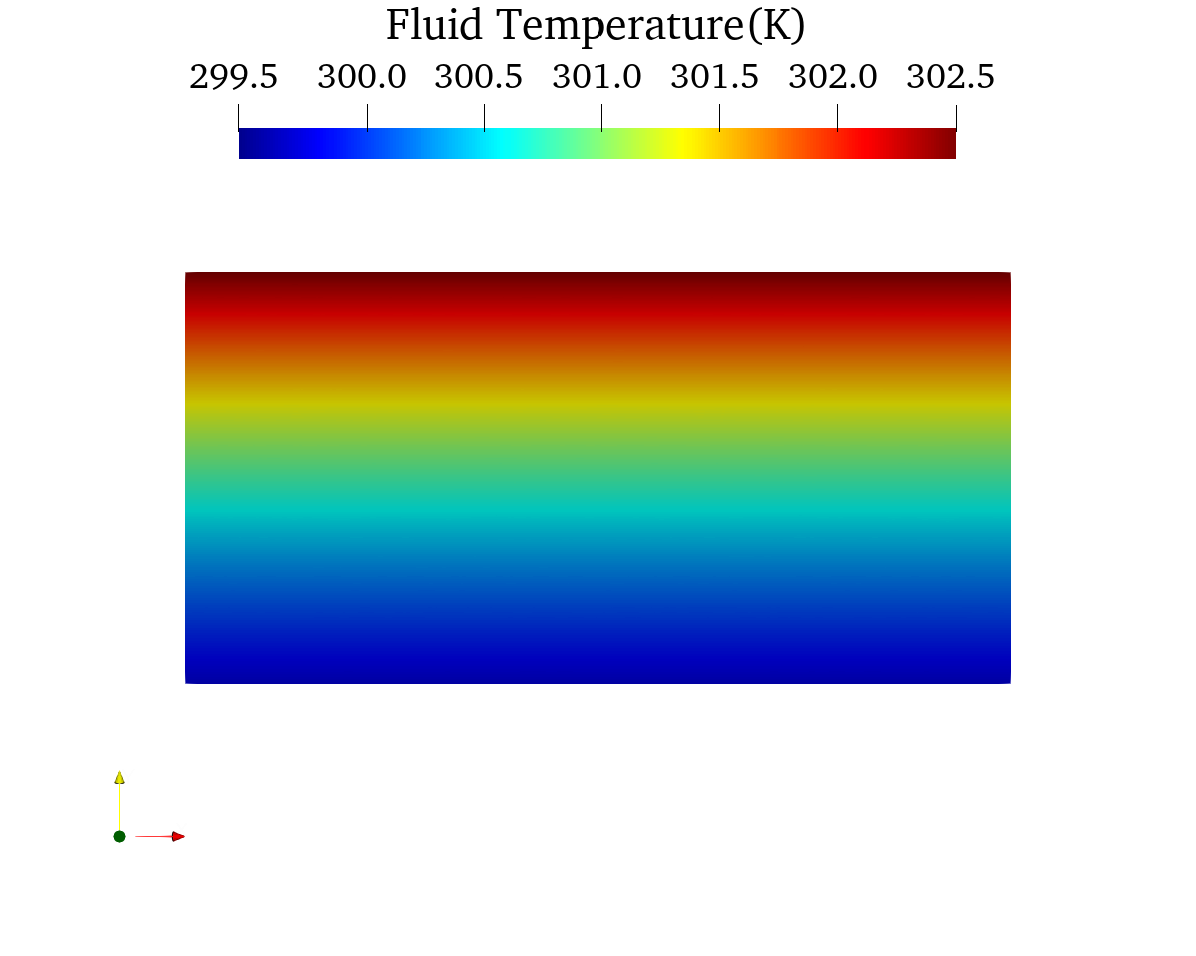}
 }  
 \vspace{-0.1in}
 \caption{
 Numerical experiments of the thermal response of silica aerogel using the two-phase heat transfer computational model \eqref{eq:thermal_bcs}.
 (a) Solid temperature $\theta_s$ and fluid temperature $\theta_s$ profiles through the sample thickness at the times $t$ = 0.1 s and $t$ = 5.0 s;
 (b,c) snapshots of the solid and fluid temperatures at time $t$ = 1.5 s.}
	\label{fig:thermal_forward}
	\vspace{-0.01in}
\end{figure}

\begin{figure}[h!]
 \centering
 \vspace{-0.1in}
 \subfloat[]{\includegraphics[width= 0.45\textwidth]{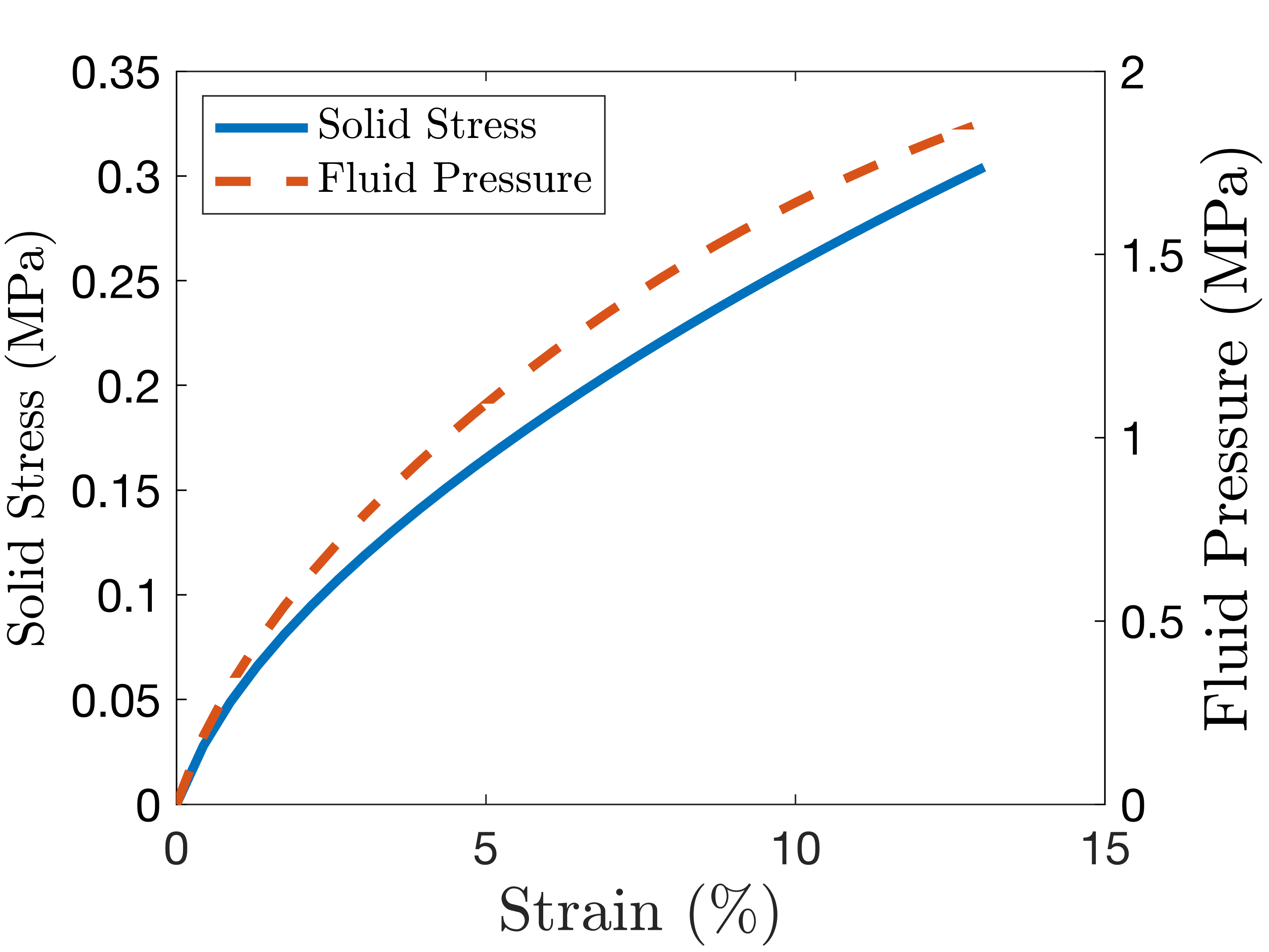}
 }
 \\\vspace{-0.1in}
 \subfloat[]{\includegraphics[trim=2.0in 3.0in 2.0in 0.0in, clip, width= 0.36\textwidth]{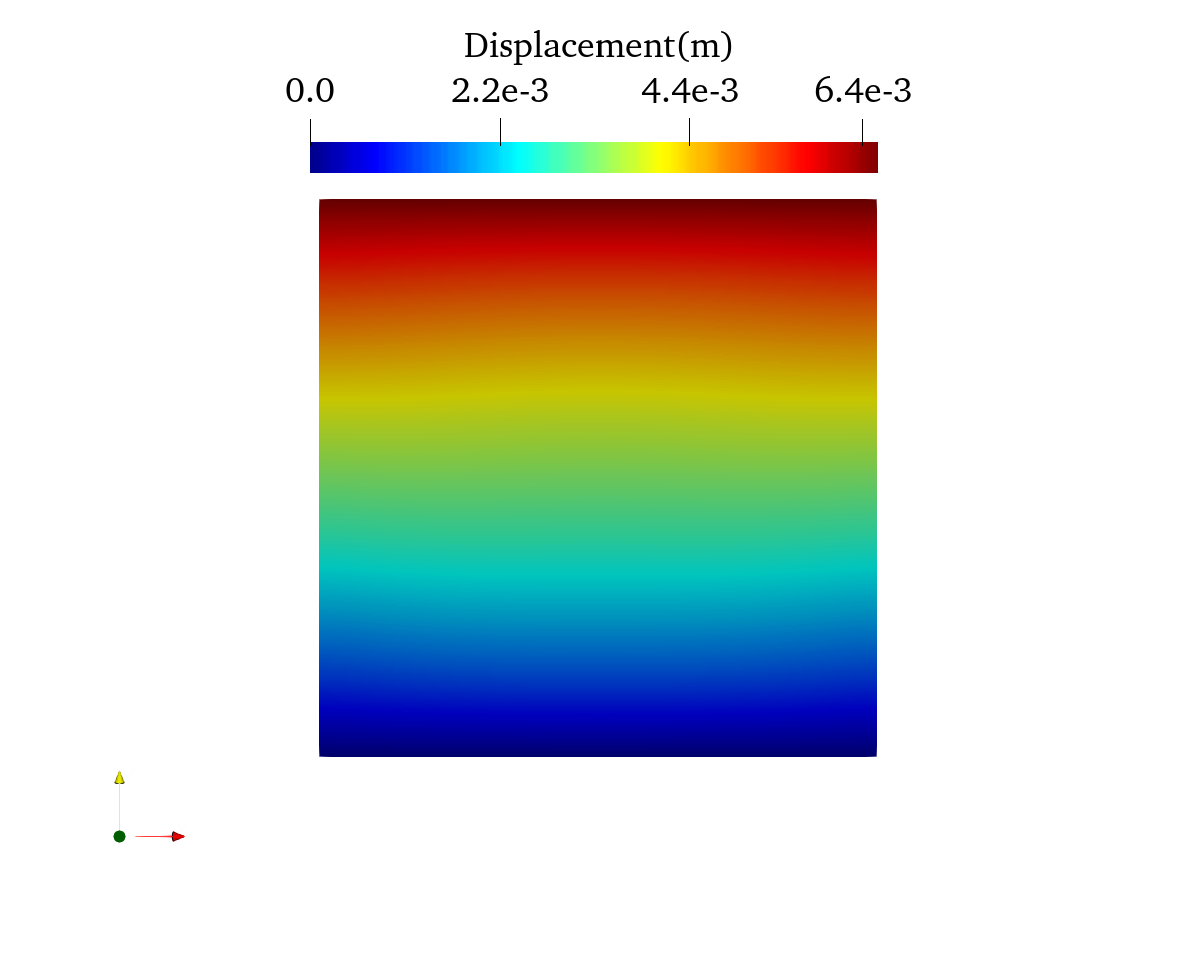}
 }
 \subfloat[]{\includegraphics[trim=2.0in 3.in 2.0in 0.0in, clip, width= 0.35\textwidth]{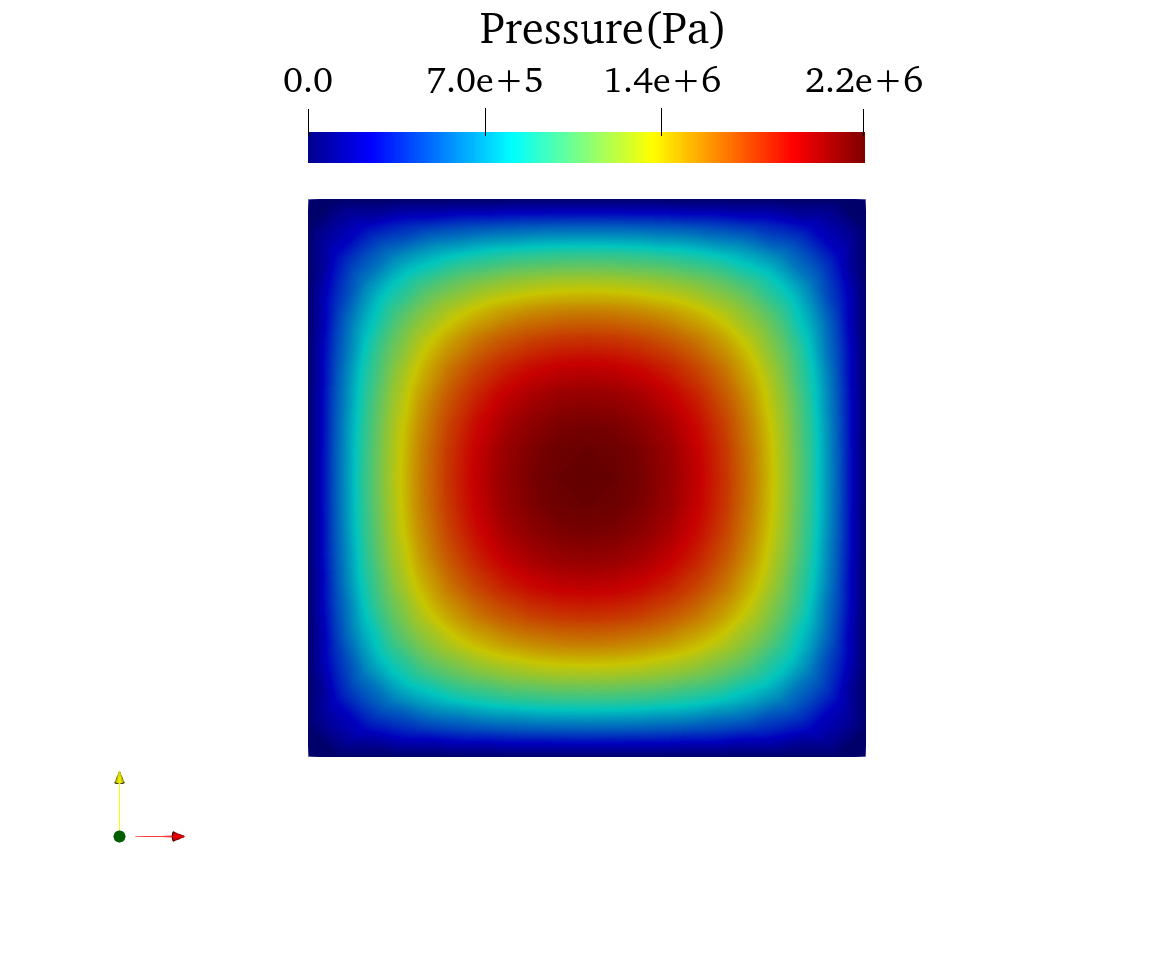}
 }  
 \vspace{-0.1in}
 \caption{
 Numerical simulation of the mechanical response of silica aerogel using the two-phase deformation computational model \eqref{eq:mech_bcs}.
 (a) variation of the solid phase stress $\mathbf{T}_s$ component along $y$-axis 
 \blue{and 
 the fluid pressure $p$ with the applied strain};
 (b) snapshots of the displacement and fluid pressure at time $t$ = 100 s.}
	\label{fig:mech_forward}
	\vspace{-0.1in}
\end{figure}

The 2D finite element solution of the heat transfer and deformation models of aerogel is implemented in an open-source computing platform, FEniCS \cite{fenics}, \blue{and verified using multiple benchmark problems, including Terzaghi's consolidation problem.}
The numerical simulation results of the heat transfer model in terms of the time evolution of solid and fluid temperature through the domain are shown in Figure \ref{fig:thermal_forward}.
For this simulation, a silica aerogel with 90\% porosity ($\phi_f=0.9, \phi_s=0.1$) and thickness of 0.6cm is considered.
The $\bar{\theta}_{hot} = 302.5 K$ is imposed at the top boundary, 
the convection to the room temperature $\bar{\theta}_{cold} = 297.5 K$ is considered at the bottom boundary,
and the model parameters are assumed as
$\kappa_s = 0.5 W/(m.K)$,
$\kappa_f = 0.08 W/(m.K)$,
$h = 60000 W/(m^{2}.K)$,
$h_{air} = 10 W/(m^{2}K)$.
Figure \ref{fig:thermal_forward} shows that, 
\blue{despite the higher conductivity of solid compared to the fluid phase 
($\kappa_s > \kappa_f$),
the fluid temperature diffuses faster in the domain compared to the solid temperature due to the higher fluid volume fraction (porosity)
of the silica aerogel.}
Figure \ref{fig:mech_forward} shows the numerical solution of the multiphase deformation model in \eqref{eq:mech_bcs}.
Similar to the uniaxial compression tests in section \ref{sec:data}, a silica aerogel sample with porosity of 90\% is 
subjected to incremental displacements with a constant strain rate of $1.36 \times 10^{-3} (1/s)$ and measurements are taken at time increments of $\Delta t=$ 3.3 s. 
The model parameters are 
$C = 8.5\times 10^{-9} ({1}/{Pa})$,
$k = 1.0\times 10^{-13} m^2$,
$E = 0.7\times 10^{6} Pa$,
$\nu = 0.3$.
The \blue{solid phase} stress-strain relationship and the fluid pressure response at the center of the sample are shown in Figure \ref{fig:mech_forward}(a), and snapshots of displacement and pressure at time $t$ = 100 s are presented in 
Figure \ref{fig:mech_forward}(b,c).
Due to the relatively low permeability of the silica aerogel, the fluid pressure does not fully dissipate during the deformation process, leading to the nonlinear stress-strain response.

\blue{
We note that performing the uncertainty analyses in Section \ref{sec:uq_analysis} using the 3D finite element model is prohibitive due to high computational costs. We thus resort to the 2D solution of the multiphase model and
leverage the noise model \eqref{eq:noise} to account for the uncertainty due to the numerical approximation as another source of modeling error (see Section \ref{sec:calibration}). 
However, to estimate the error incurred by the 2D assumption,
we conducted 3D analyses of the results presented in Figures \ref{fig:thermal_forward} and \ref{fig:mech_forward} with the same parameter values and discretization resolution. 
Comparing the 2D and 3D results indicates that the primary source of error is the computed pore pressure, while the errors in other outputs are negligible. In particular, the 2D model under-predicted the pressure dissipation compared to the 3D model. The average errors in solid stress and fluid pressure are 0.7\% and 23\%, respectively, while the errors in 2D and 3D simulations of solid and fluid temperatures are below 0.0017\%.
}

\subsection{Uncertainty analyses of the thermomechanical multiphase model}\label{sec:uq_analysis}

\begin{figure}[h!]
 \centering
 \vspace{-0.1in}
	\includegraphics[width= \textwidth]{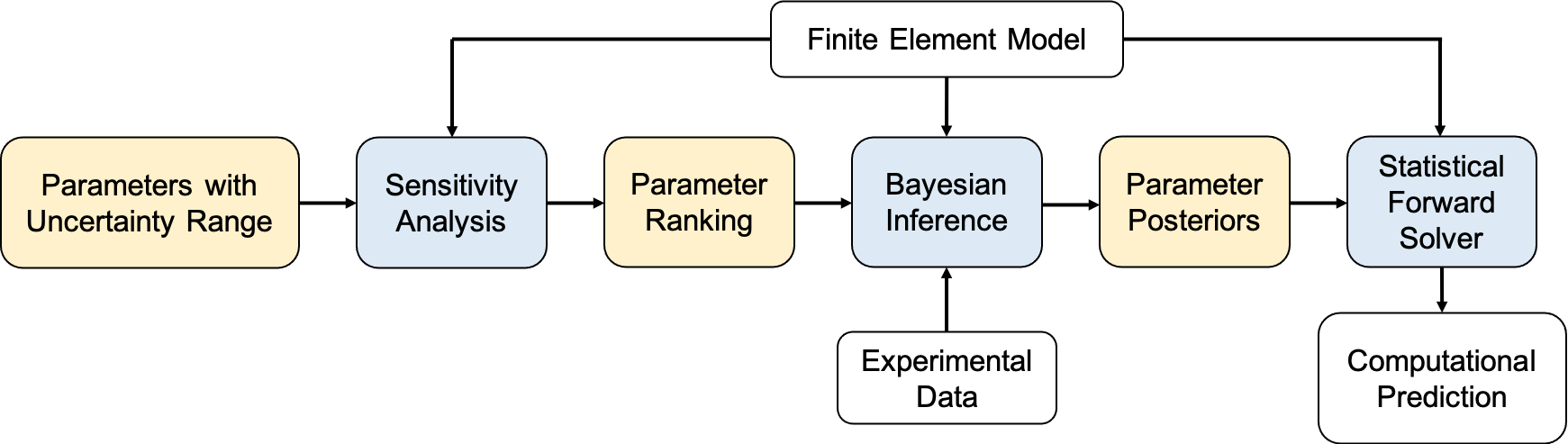}
 \vspace{-0.2in}
 \caption{
 \blue{
Flowchart of the framework implemented in this work for predictive computational modeling with quantified uncertainty.
It includes the sensitivity analysis for model parameter ranking,
Bayesian calibration for determining the probability distribution of parameters from measurement data and a statistical forward solver for computational prediction while assessing the level of confidence in the predicted QoI.
 }}
	\label{fig:flowchart}
	\vspace{-0.1in}
\end{figure}

\blue{
To characterize the uncertainty in the prediction of the aerogel multiphase model, we implement a general UQ framework leveraging the methods described in Section \ref{sec:uq}. Figure \ref{fig:flowchart} shows the flowchart of this framework that systematically characterize the uncertainty and assess the reliability of computational predictions delivered by physics-based models. 
Starting from a set of uncertain model parameters, global sensitivity analysis is conducted to rank the model parameters based on their contributions to the variability of the model outputs.
The sensitivity analysis guides dimension reduction of the parameter space used for calibration, such that unimportant parameters are assumed to be constant or deterministic.
The Bayesian inference is then conducted to learn the parameters from measurements, i.e., updating the parameters priors to the posterior probability distributions.
The calibrated model with uncertain parameters enables predicting possibly more complex physical systems for which observational data are not available. To this end, using the parameter posteriors, the statistical forward problem is solved to propagate the uncertainty from the model parameters to the computed quantity of interest (QoI).
The rest of this section describes the implementation of this framework for predictive computational modeling of silica aerogel's thermal and mechanical performances. 
}

\subsubsection{Parameter sensitivity analysis}\label{sec:sa}

The sensitivity scenarios of the multiphase heat transfer and deformation models are the domain and boundary conditions, which are chosen according to the experimental measurements described in Section \ref{sec:fe}.
\blue{Two model outputs are chosen in the parameter sensitivity analysis: for the heat transfer model, $Q^T$ is the mixture heat flux through the bottom surface in as depicted in Figure \ref{fig:BC}(a); for the deformation model, $Q^M$ is the total strain energy over the applied deformation. Specifically, }

\begin{eqnarray}
Q^T &=&  
|\hat{q}|,
 \label{eq:QoI_T}
\\
Q^M &=& \int_{\Omega^M}
( \mathbf{T} : \mathbf{E} )
 d\mathbf{x}, 
 \label{eq:QoI_M}
\end{eqnarray}
\blue{where 
$\mathbf{E}$ is total strain corresponding to the applied $\bar{u}$ in Figure \ref{fig:BC}(b).
Note that both model outputs are the quantities that will be used for Bayesian calibration. 
Thus, the parameter sensitivity will possibly provide information to aid the calibration process.}

\begin{table}[h!]
 \label{table:SAparameters}
 \footnotesize
 \centering
 \begin{tabular}{c c c c c} 
 \hline
 \hline
 \textbf{Parameter} & \textbf{Physical meaning} & \textbf{Unit} & \textbf{Uncertainty Range} & \textbf{Prior PDF} \\ [0.5ex] 
 \hline
 \(\kappa_{s}\) & solid conductivity & $W/(mK)$ & $[0.1,1.4]$ & $\mathcal{U}(0.1,1.4)$\\ 
  
 \(\kappa_{f}\) & fluid conductivity & $W/(mK)$ & $[0.02,0.14] $ & $\mathcal{U}(0.02,0.14) $\\
  
 \(h\) & interstitial coefficient & $W/(m^{2}K)$ & $[1.0,100000.0]$ & $\mathcal{U}(1.0,100000.0)$\\
  
 \(m^T\) & thermal noise multiplier & -- &-- & $\Gamma^{-1}(102,103)$ \\  
 \hline
 \(E\) & elastic modulus & $Pa$ & $[1000.0, 20000.0]$ & $\mathcal{U}(1000.0, 20000.0)$ \\ 
  
 \(\nu\) & Poisson's ratio & -- & $[0.25,0.4]$ & fixed at 0.3 \\

 \(\log(k)\) & log-permeability coefficient & $\log(m^2)$ & $[-54.54, -14.54]$ & $\mathcal{N}(-34.54, 10.0)$\\
  
 \(\log(C)\) & log-compressibility of fluid & $\log(1/Pa)$ & $[-34.72, -0.672]$ & $\mathcal{N}(-20.72, 7.0)$\\
  
 \(m^M\) & mechanical noise multiplier & -- & -- & $\Gamma^{-1}(102,103)$ \\  
 \hline 
 \hline  
 \end{tabular}
 \caption{
 \blue{
The model parameters are the two-phase heat transfer and deformation models of silica aerogel. 
Uncertainty Range is the initial range of the parameters taken into account for the variance-based global sensitivity analyses.
Probability distribution functions (PDF) of parameters is the priors of the Bayesian training, where $\mathcal{U}(\cdot,\cdot)$ indicates the uniform probability distribution, $\mathcal{N}(\cdot,\cdot)$ indicates the Gaussian probability distribution, and $\Gamma^{-1}(\cdot,\cdot)$ indicates the inverse gamma probability distribution.
}
 }
\end{table}

\begin{figure}[h!]
 \centering
 \subfloat[]{\includegraphics[width= 2in]{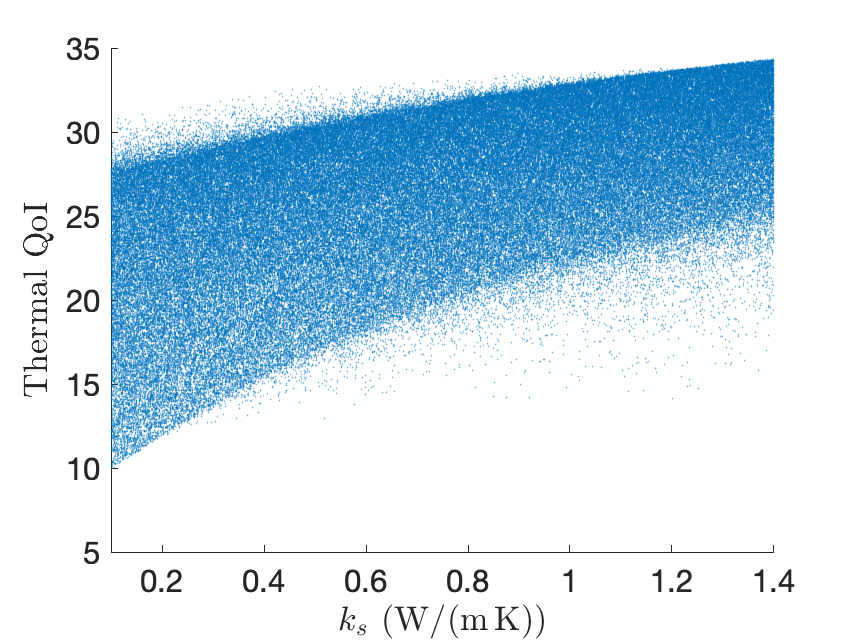}}
 \subfloat[]{\includegraphics[width= 2in]{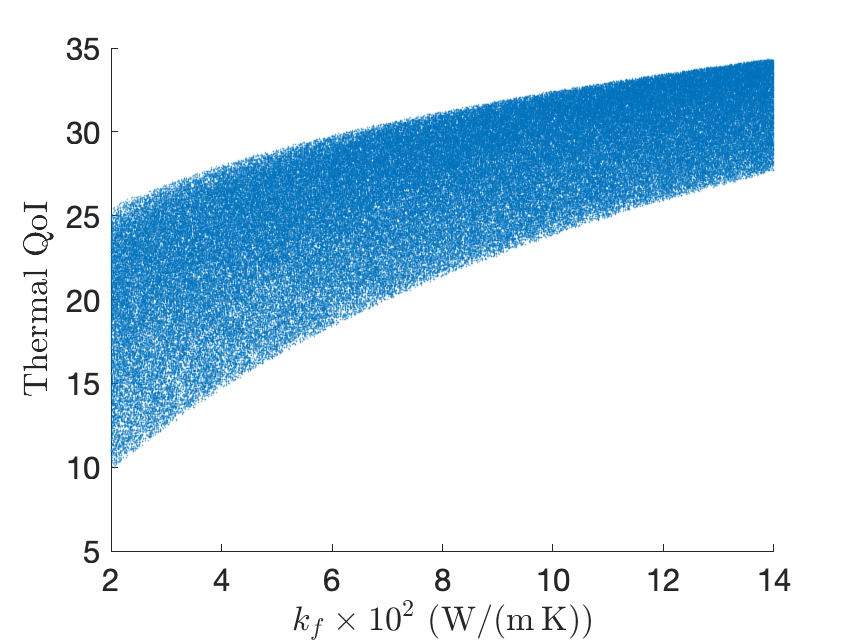}}
 \subfloat[]{\includegraphics[width= 2in]{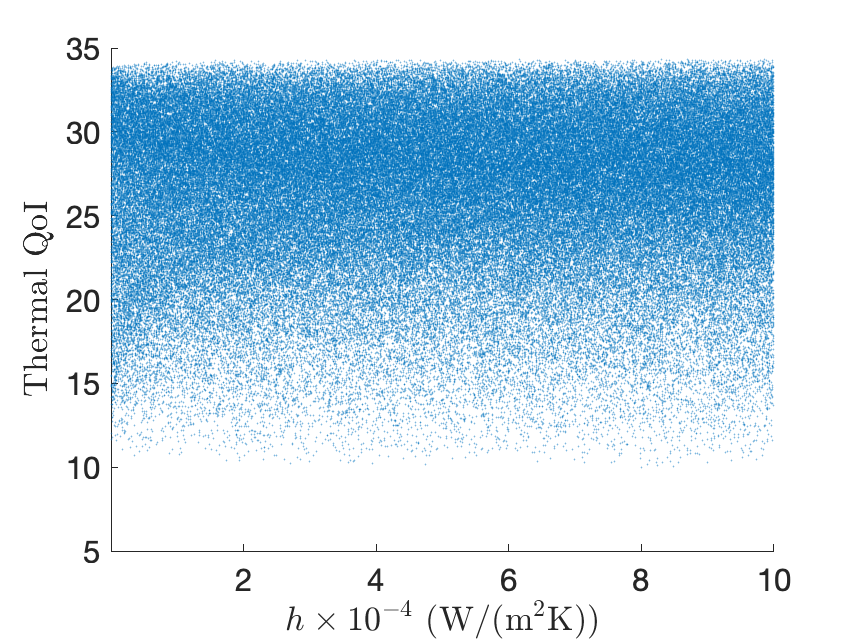}}
  
 \subfloat[]{\includegraphics[width= 2in]{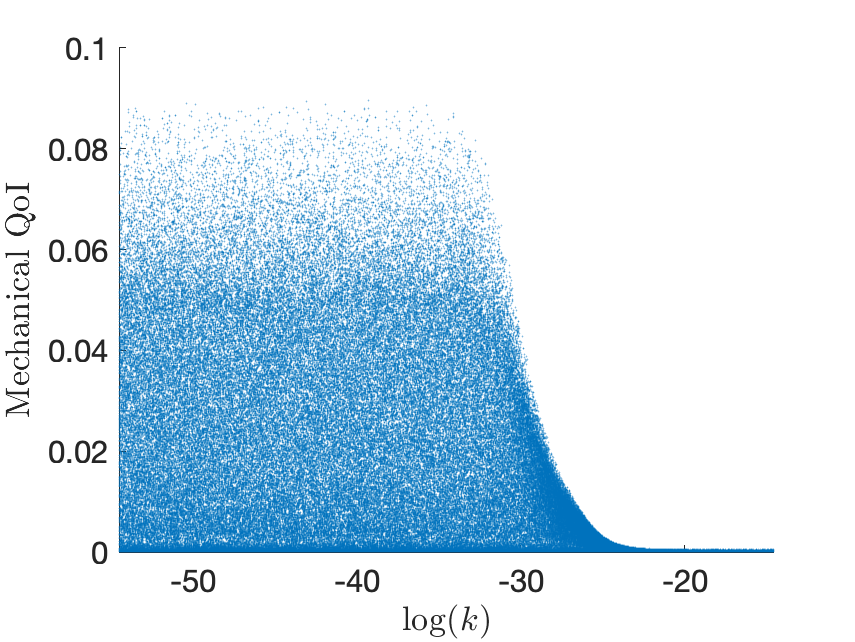}}
 \subfloat[]{\includegraphics[width= 2in]{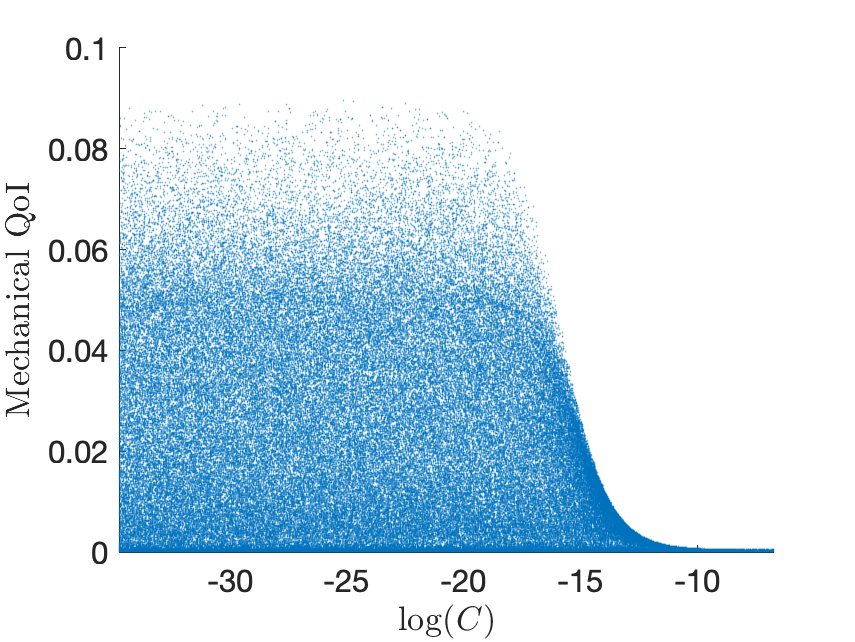}}
 
 \subfloat[]{\includegraphics[width= 2in]{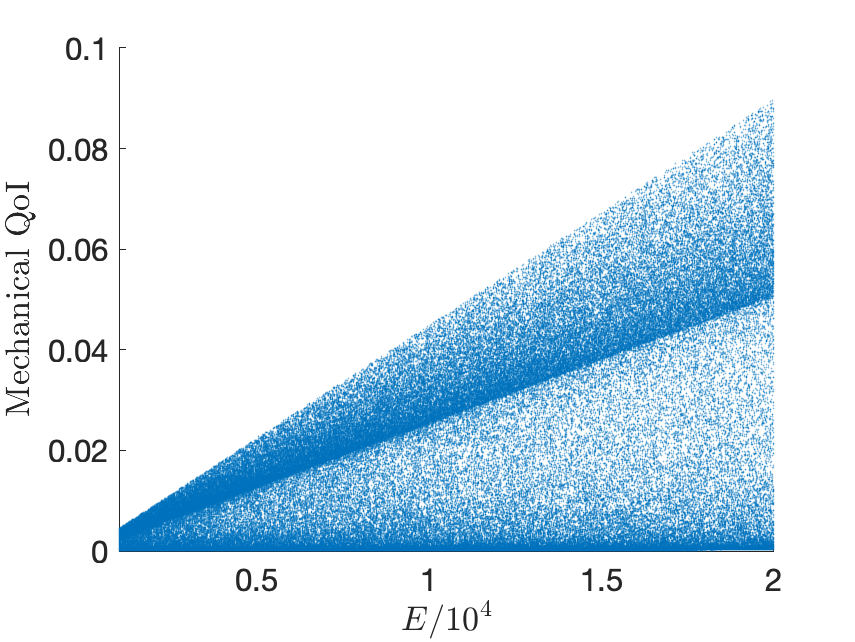}}
 \subfloat[]{\includegraphics[width= 2in]{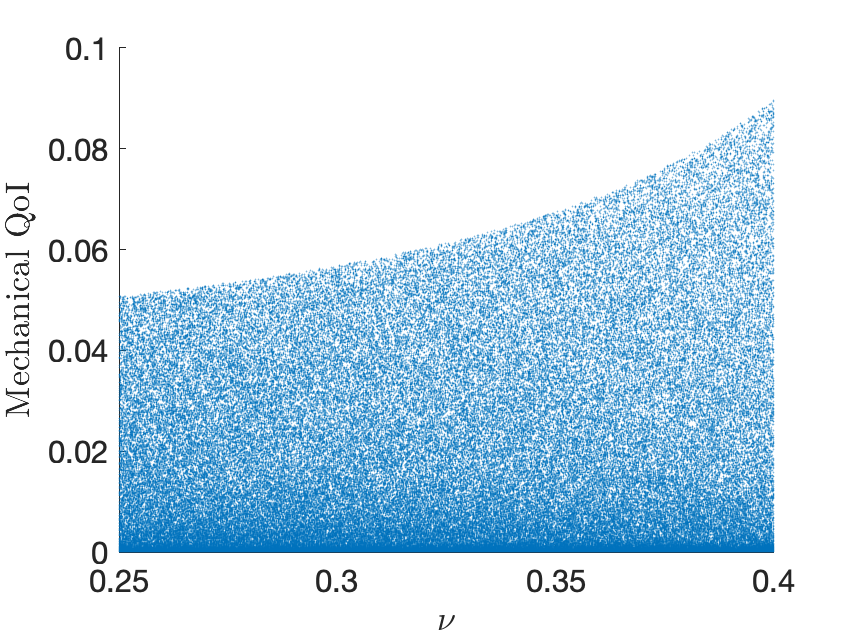}}
 \caption{
 \blue{
 Scatter-plots of the two-phase heat transfer and deformation model parameters with the thermal and mechanical model outputs.
 The results of silica aerogel with
 (a-c) 90\% of porosity ($\phi_f = 0.9$) for the heat transfer model;
 (d-g) 10\% of porosity ($\phi_f = 0.1$) for the deformation model.
 }
 }
 \label{fig:scattter}
\end{figure}

For visual assessment of parameter sensitivity, we use scatter-plots \cite{saltelli2008}, which are plots of the model output versus random samples drawn from the uncertain input parameters. The parameters with a significant sensitivity on the model output present a distinctive pattern in the scatter-plot. 
\blue{The scatter-plots of the heat transfer model are shown in Figure \ref{fig:scattter} (a-c), and those of the deformation model are shown in Figure \ref{fig:scattter} (d-g), both for the silica aerogel with a porosity of 90\% ($\phi_f = 0.9$). }
\blue{The thermal model output $Q^T$ is computed using 200000 uniform samples via Latin Hypercube sampling. For the mechanical model, 240000 samples are used to accommodate a larger number of model parameters.}
\blue{
It is seen from the scatter-plots that the solid and fluid conductivities, $\kappa_s$ and $\kappa_f$, are both critical contributors to the variance of thermal model output for the silica aerogel, as they are exhibiting patterns of the clouds in the scatter-plots (Figure \ref{fig:scattter}(a-c)).
The scatter-plots of the deformation model in Figure \ref{fig:scattter} (d-g) indicate that all four parameters contribute to the output uncertainty.
Interestingly, the permeability $k$ and the fluid compressibility $C$ show unique patterns of the clouds in the scatter-plots suggesting a very strong sensitivity over specific ranges of these parameters, i.e., 
$-35< \log(k)<-22$ and
$-18< \log(C)<-10$,
while the model output is insensitive to the variations of $C$ and $k$ outside those ranges. 
As it is discussed in Section \ref{sec:calibration}, we leverage this information to construct informative priors for Bayesian inference.
}

\begin{figure}[h!]
 \centering
 \vspace{-0.2in}
 \subfloat[]{\includegraphics[width= 0.48\textwidth]{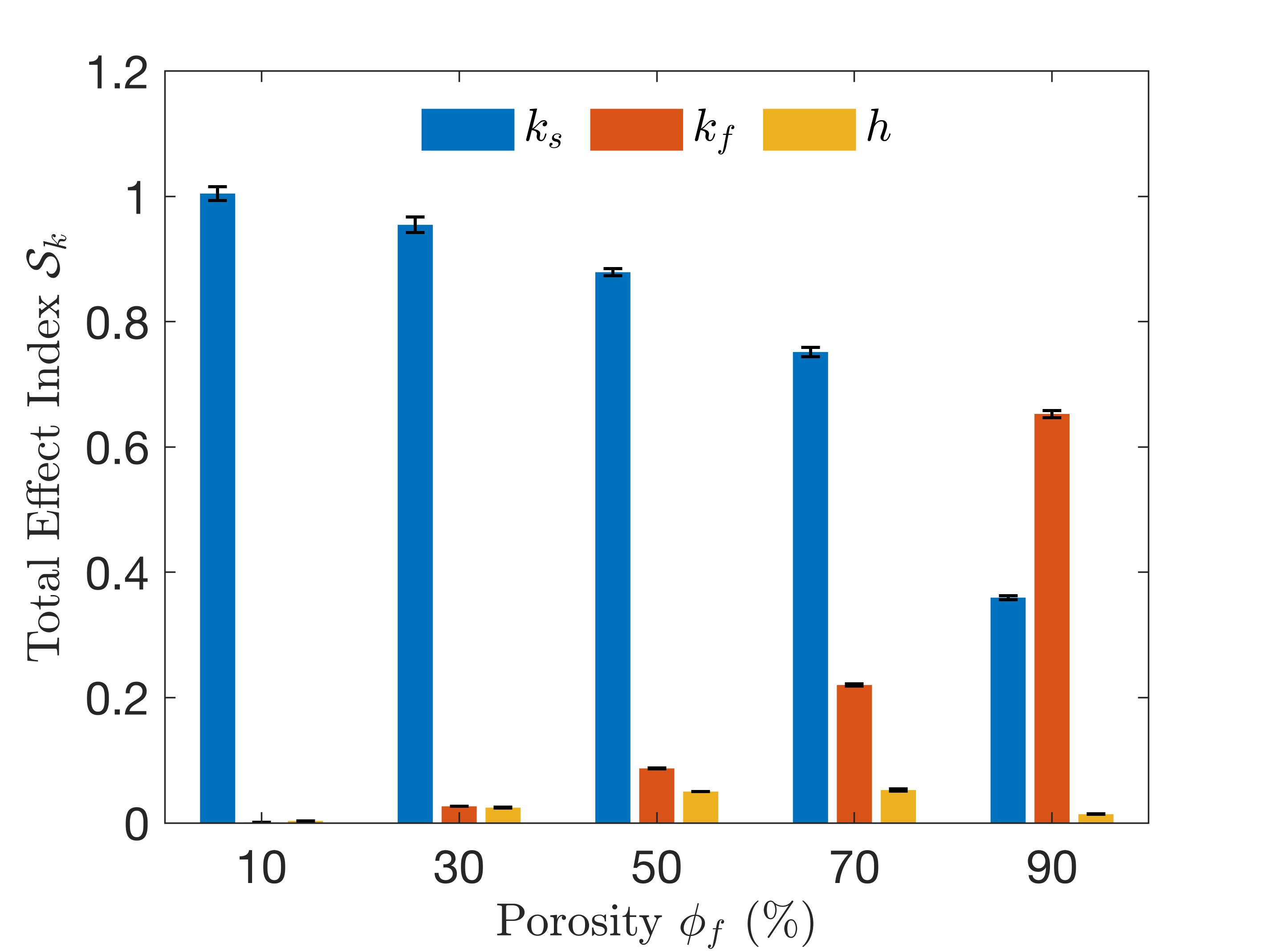}}
 \subfloat[]{\includegraphics[width= 0.48\textwidth]{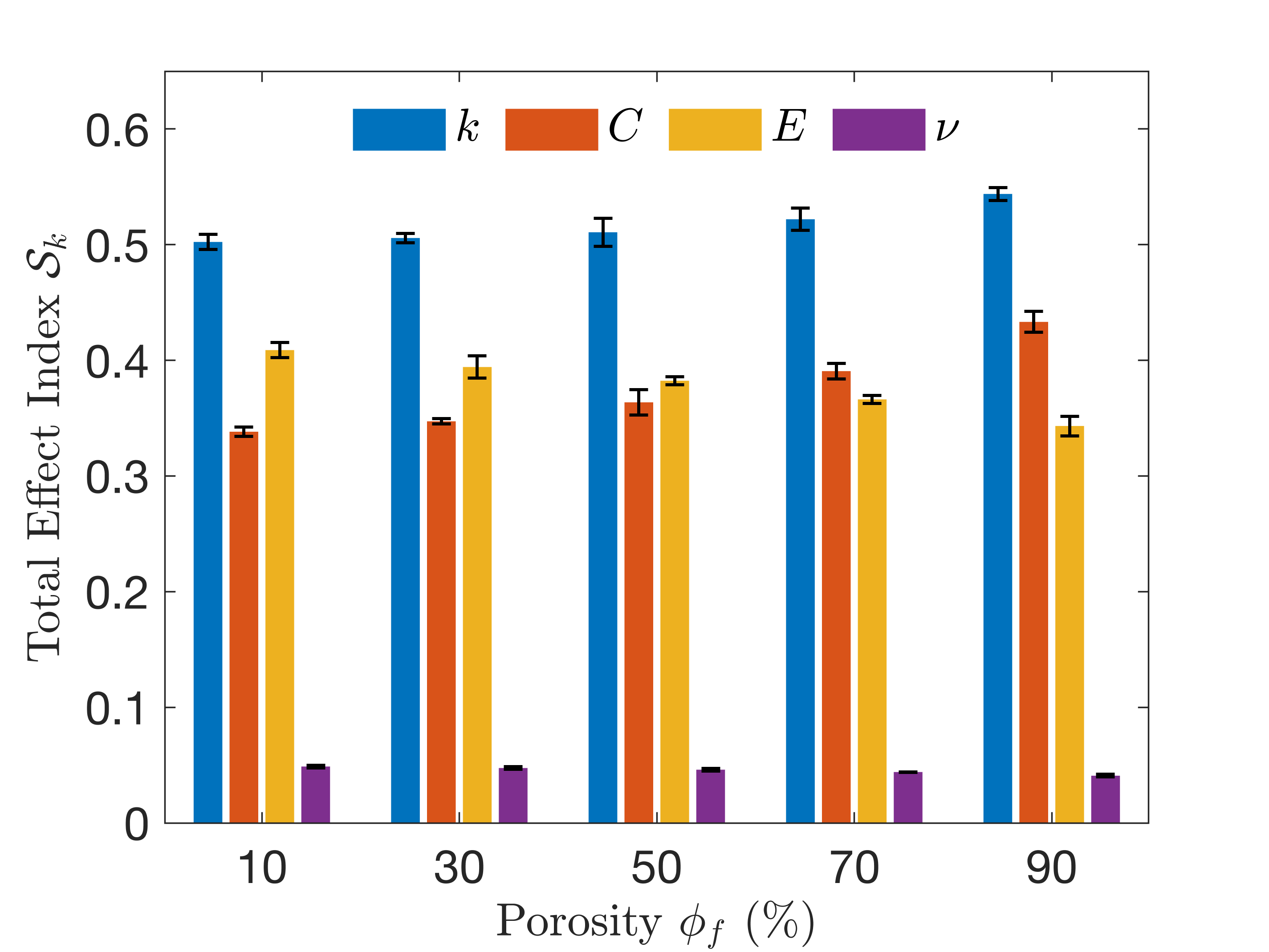}}
 \vspace{-0.1in}
 \caption{
 \blue{
 The results of variance-based global sensitivity analyses of the aerogel's heat transfer and mechanical models with different porosities $\phi_f$.
 The total effect sensitivity indices $\mathcal{S}_k$ for
 (a) the heat transfer model parameters;
 (b) the deformation model parameters.
 }
 }
 \label{fig:sensitivity}
\end{figure}

A quantitative \blue{method for ranking of the model parameters} is the VSA described in Section \ref{sec:sensitivity}. 
The \blue{range} of the parameters used for the variance-based sensitivity analyses are shown in \blue{the fourth column of} Table 1.
$N$=10000 samples are drawn for the parameter distributions using Latin Hypercube sampling to compute the total-effect sensitivity indices $\mathcal{S}_k$, using the estimator \eqref{eq:mc_sensitivity}.
The indices are computed for different porosities to understand the effect of the fluid volume fraction $\phi_f$ on parameters' importance. 
\blue{Since this Monte Carlo estimator is stochastic, five independent calculations are performed for each porosity to obtain means and standard deviations for the indices}, and
the results are shown in Figure \ref{fig:sensitivity}.
The parameter sensitivity of the heat transfer model in Figure \ref{fig:sensitivity}(a) shows that by increasing the aerogel porosity, the contribution of the solid phase conductivity $\kappa_s$ \blue{to the variance of the heat flux $Q^T$ decreases,
while the sensitivity indices of the fluid conductivity $\kappa_f$ increases.}
Therefore, the parameter \blue{$\kappa_f$} is the most \blue{important parameter in forward uncertainty propagation through the heat transfer model} for the silica aerogels with $\phi_f = 0.84-0.94$, \blue{corresponding to} the experimental measurements in Figure \ref{fig:exp_data}(a).
\blue{As expected, the monotonic increase of sensitivity index of $\kappa_s$ and the monotonic decrease of sensitivity index of $\kappa_f$ with aerogel porosity indicates that the volume fraction of each phase determines their contribution to the heat transfer simulations. Furthermore, when both phases are equally present in the binary mixture ($50\%$ porosity), the interstitial coefficient $h$ is highest, providing the most substantial heat exchange between the constituents.}
\blue{The total-effect sensitivity indices of the deformation model in Figure \ref{fig:sensitivity}(b) indicate that the permeability $k$ is the most influential parameter on the model outputs' variance. 
Moreover, the sensitivity indices of the fluid mass balance parameters, compressibility $C$ and permeability $k$, increase as the aerogel porosity $\phi_f$ increases.
The solid phase Young's modulus of aerogel $E$ becomes less influential in higher porosity due to higher fluid pressure development. }
The variation of the total-effect sensitivity indices with the aerogel porosity is controlled by the fluid pressure development and dissipation in the domain. As the aerogel porosity increases, the fluid pressure $p$ increases, resulting in higher fluid strain energy than the solid phase strain energy in the silica aerogels with $\phi_f>0.8$.

Given the low computational costs of performing the global sensitivity analyses, compared to Bayesian inference, the VSA results can be used for model reduction, i.e., reducing the space of the uncertain parameters. 
In particular, the parameters with sufficiently low sensitivity indices can be taken as deterministic or fixed values, lowering the computational cost of solving the statistical inverse problem. 
According to the results presented in Figure \ref{fig:sensitivity}, we consider a fixed value of the Poisson's ratio $\nu= 0.3$ for the deformation model calibration and predictions, as this parameter does not have a substantial effect on the mechanical model output.

\subsubsection{Bayesian inference}\label{sec:calibration}

Guided by the parameter sensitivity analyses, we conduct Bayesian inference to learn the parameters of the two-phase models from the experimental measurements of silica aerogel presented in Section \ref{sec:data}. 
\blue{The prior distributions of the parameters are tabulated in the last column of Table 1.
Due to the lack of prior information about the multiphase thermal model for the new silica aerogel material, we consider non-informative priors for the heat transfer model parameters. However, we leverage the unique patterns observed in the scatter-plots of Figure \ref{fig:scattter} (d,e) to construct prior distributions of the deformation model parameters, fluid permeability $k$, and compressibility $C$.
In particular, we imposed higher probabilities of these parameters within the range they exhibit strong sensitivity to the model output via assigning Gaussian prior distributions to $\log(C)$ and $\log(k)$.
}

\begin{figure}[h!]
 \centering  
 \vspace{-0.2in}
 \subfloat[]{\includegraphics[width= 0.48\textwidth]{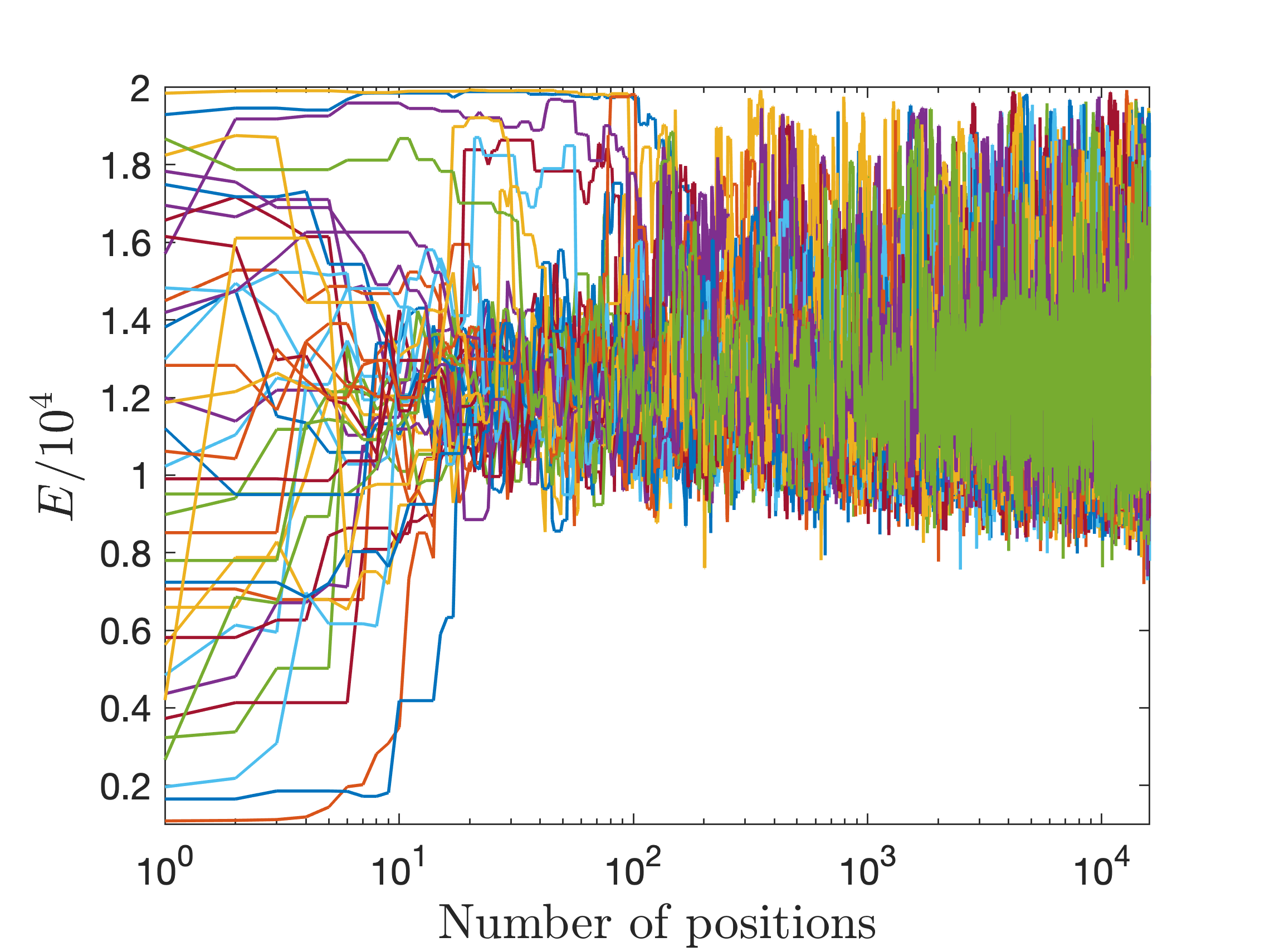}}
 \subfloat[]{\includegraphics[width= 0.48\textwidth]{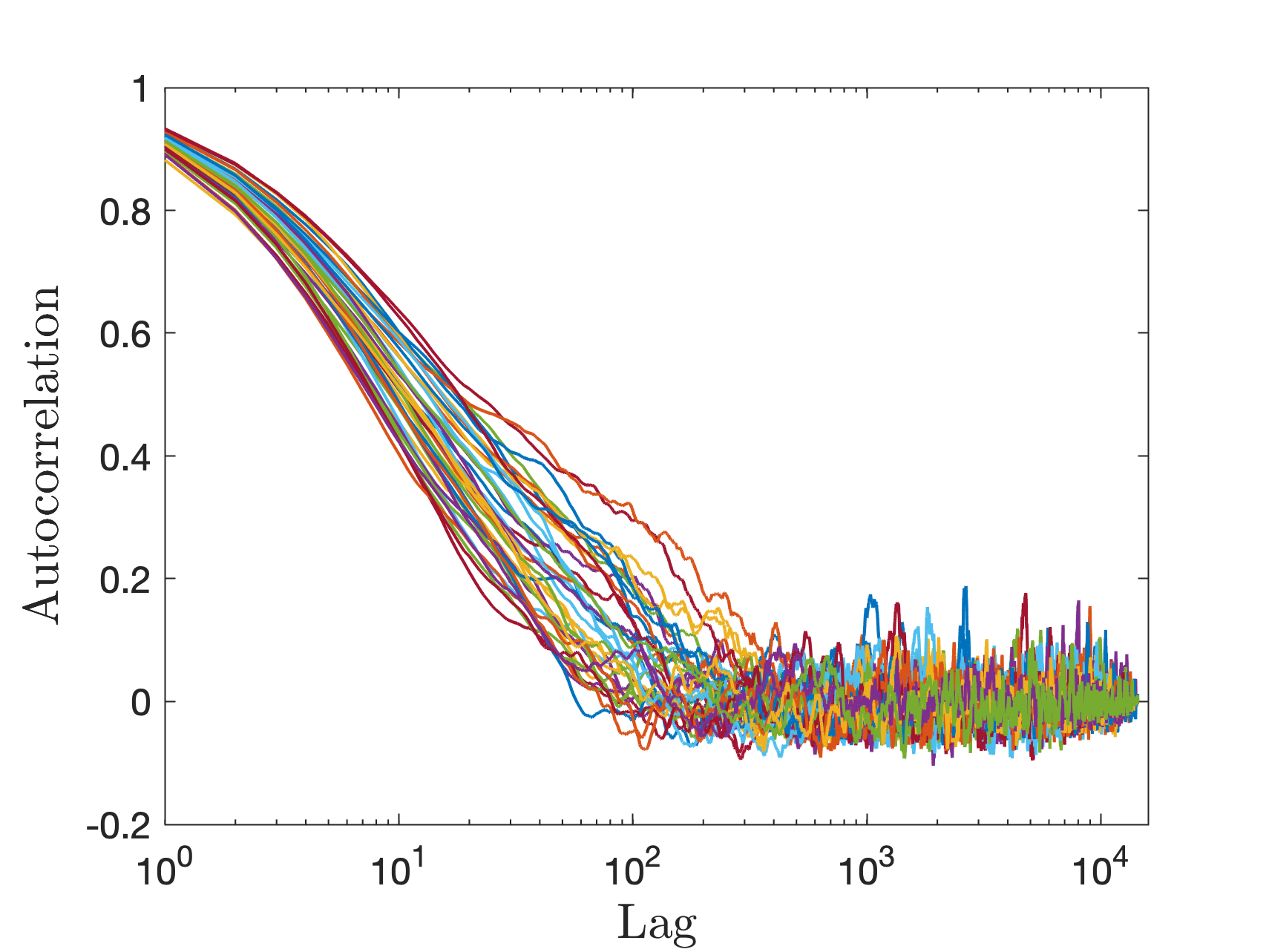}}
 \vspace{-0.1in}
 \caption{
 \blue{
 Diagnostics plots of the {40} Markov chains with the length of {16000} for the Bayesian calibration of the parameter 
 $E$. 
 The first 10\% of points in these plots, exhibiting significant auto-correlation, are excluded from the final posteriors as they represent the initial burn-in period. 
 (a) Trace plot of the chains indicating good mixing properties of the samples;
 (b) Auto-Correlation Function (ACF) for the samples, where the Lag indicates the number of steps between samples. Trace and ACF plots of the other parameters exhibit similar behavior to those observed in these plots. 
}
}
 \label{fig:trace}
\end{figure}

The DRAM algorithm, implemented in the parallel object-oriented statistical library DAKOTA \cite{dakota2020}, is then employed to solve the Bayesian inference (\ref{eq:bayes}) \blue{on the finite element model}. 
To reduce the high computational burden of the inherently sequential MCMC algorithms, we made use of a parallel workload, wherein each chain is run on a computing processor, while each forward model solution is performed in parallel and on multiple processors.
\blue{In particular, 40} chains are used for calibrating each (heat transfer and deformation) model with chain lengths of \blue{16000}. Each chain was initialized from different parameter values sampled from their corresponding priors. The initial 10\% samples of each chain are discarded (burn-in period) to ensure the chain has reached a state sufficiently close to the stationary distribution.
Figure \ref{fig:trace} shows example diagnostics plots of the Markov chains for inferring the parameter $E$.
The trace plot of the \blue{40} chains in Figure \ref{fig:trace}(a) shows that chains starting from different values mix well after 100 steps. The autocorrelation function (ACF) of the MCMC sequence in Figure \ref{fig:trace}(b) shows a rapid decay indicating that samples at a relatively small lag from each other can be considered as independent.

\begin{figure}[h!]
 \centering  
 \includegraphics[trim=0.0in 0.15in 0in 0.15in, clip, width= 0.8\textwidth]{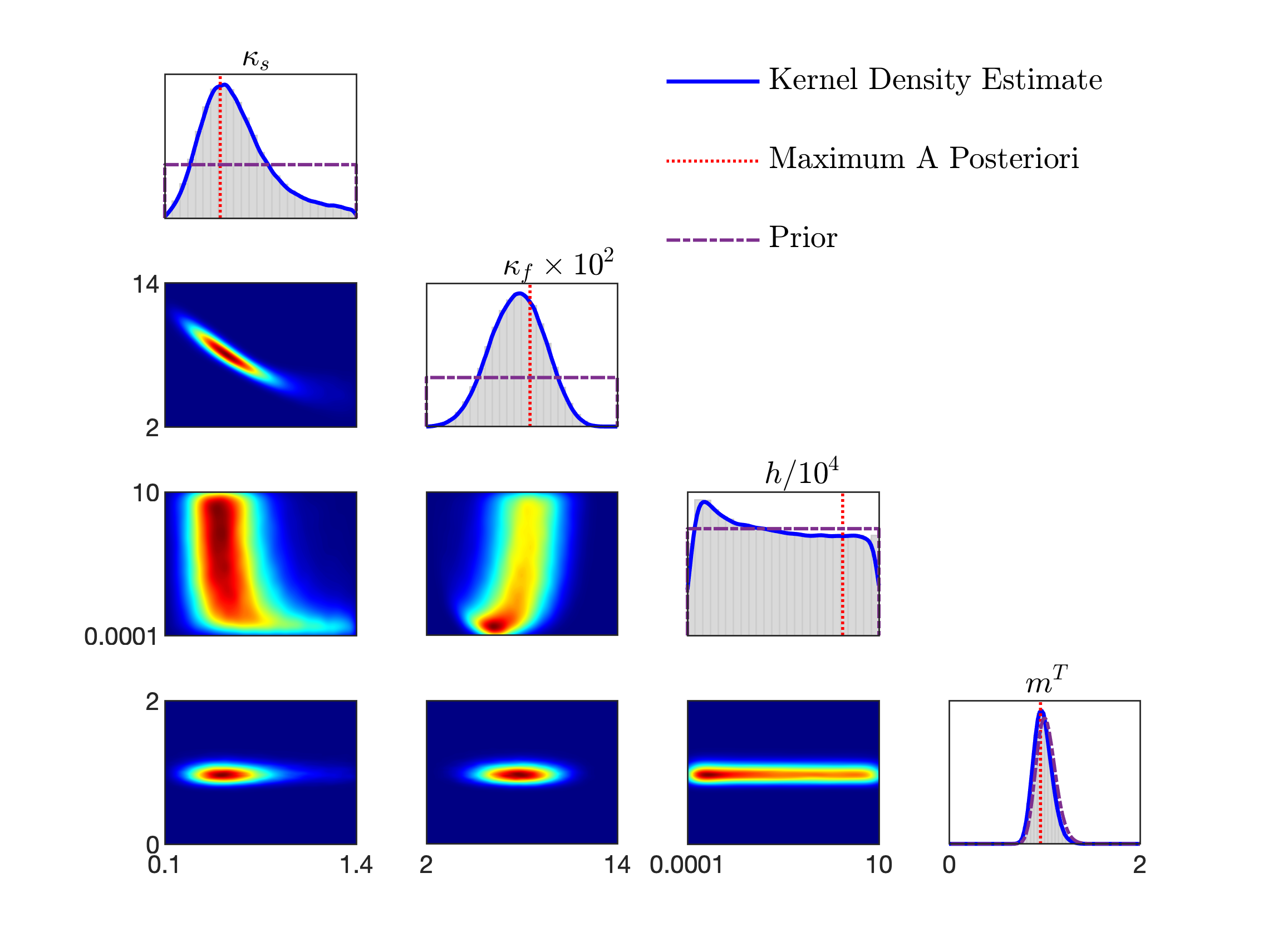}
 \vspace{-0.15in}
 \caption{
 \blue{
 The multiphase heat transfer parameter posteriors were obtained from the Bayesian inference. 
 The training data sets consist of the measured heat flux for the aerogel samples with different porosities. 
 The 1D plots represent the kernel density estimations (KDEs) of 
 the marginal posterior distributions and 
 the joint bivariate posteriors distributions are the 2D KDE plots. 
The dash-dotted lines are the priors defined in Table 1, 
 and the dashed lines are the MAP estimates in {Table 2}.
 }}
 \label{fig:thermal_posteriors}
\end{figure}

The kernel density estimations (KDEs) of the heat transfer parameter posteriors are shown in Figure \ref{fig:thermal_posteriors}, including the physical parameters $(\kappa_s, \kappa_f, h)$ and the noise multiplier hyper-parameter $m^T$ in \eqref{eq:noise}.
The training data $\boldsymbol{D}$ consists of the heat flux measurements of the silica aerogel sample shown in Figure \ref{fig:exp_data}(a) for different values of aerogel porosity. The model output $\boldsymbol{d}(\boldsymbol{\theta})$ in the likelihood function \eqref{eq:likelihood} is the corresponding heat flux of the mixture computed from the two-phase heat transfer model \eqref{eq:thermal_bcs}.
We note that measurements are from aerogel samples with different pore sizes without accounting for the experimental variability of different samples. 
\blue{
Since the two-phase heat transfer model does not capture the effect of aerogel pore size on the thermal behavior, the measurements of different pore sizes are accounted for as the microstructure uncertainty in data for the Bayesian inference.}
Figure \ref{fig:thermal_posteriors} shows both marginal distributions, corresponding to a single parameter and joint bivariate distributions of two parameters.
\blue{The posterior variances indicate the level of confidence on the calibrated parameters in the presence of data noise and modeling error}. 
In addition, the MAP point estimates, $\boldsymbol{\theta}^{MAP}$, according to \eqref{eq:MAP}, are shown in these plots \blue{that corresponds to deterministic calibration in which a single value for each parameter is identified.
Contrariwise, characterizing uncertainty through the posterior probability distributions of parameters enables assessing the reliability of prediction delivered by the calibrated model (see Section \ref{sec:prediction}). 
}  
Some of the parameters' MAP points are different from the maximum of the marginal distributions since the two-point estimates are identical only when the posteriors are normal distributions.
The posterior plots indicate that $\kappa_s$ is learned better from the data leading to the smaller posterior variance of this parameter.
On the contrary, the interstitial coefficient $h$ is not well informed by the data as judged by its wide posterior distribution.
The joint bivariate distributions of $\kappa_s$ and $\kappa_f$ indicate a strong correlation among them.
Additionally, the posterior distribution of the noise multiplier $m^T$ has the MAP value of {0.96} according to \blue{Table 2}. The close to one value of the noise multiplier indicates that, in this case, the average data variance over all the data points is sufficient to balance the trade-off between data misfit in the likelihood function and the parameter priors.
We emphasize that the thermal data variance $\sigma_i, i=1,2,\cdots, 11$ in \eqref{eq:noise} arises from the marginalization of the pore size effect ignored by the model, and thus it is the principal contributor to the model inadequacy.

\begin{figure}[h!]
 \centering  
 \includegraphics[trim=0.0in 0.3in 0in 0in, clip, width= 0.8\textwidth]{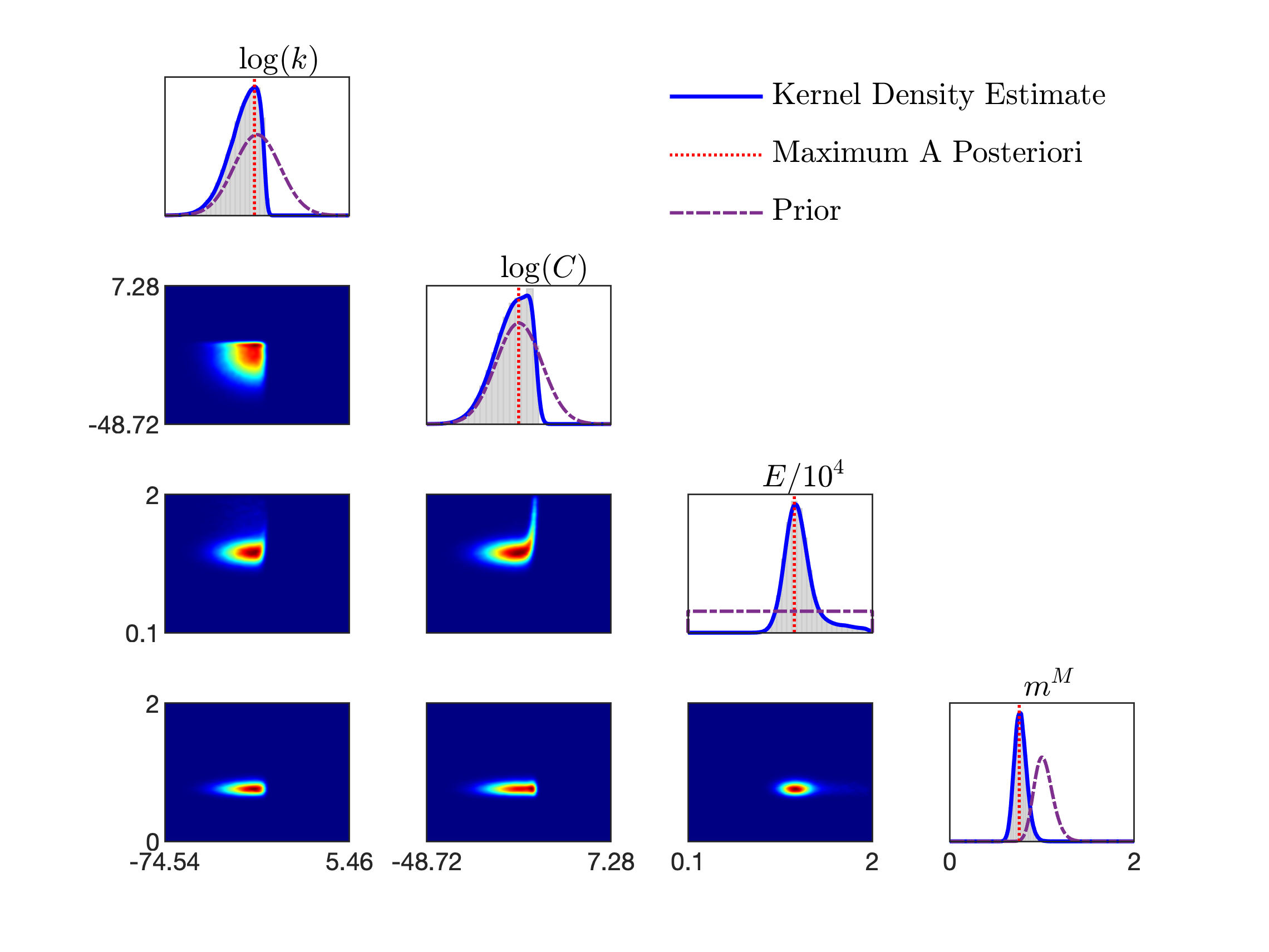}
 \vspace{-0.15in}
 \caption{
 \blue{
 The multiphase mechanical parameter posteriors were obtained from the Bayesian inference. 
 The training data sets consist of the measured stress-strain, the 3D printed aerogel sample. 
 The 1D plots represent the kernel density estimations (KDEs) of the marginal posterior distributions, and the joint bivariate posteriors distributions are the 2D KDE plots. 
 The dash-dotted lines indicate the prior distributions defined in Table 1, 
 and the dashed lines are the MAP estimates presented in Table 2.}
 }
 \label{fig:mechanical_posteriors}
 \vspace{-0.05in}
\end{figure}

Figure \ref{fig:mechanical_posteriors} presents the Bayesian inference results of the deformation model \eqref{eq:mech_bcs}, using the experimental data of Figure \ref{fig:exp_data}(b).
This figure indicates that
while the solid phase's elastic modulus is well-informed, 
\blue{the posterior parameters $C$ and $k$ are shifted toward larger values compared to the priors with wide posterior variances. }
One can thus conclude that a set of stress-strain measurements does not provide sufficient information for training the two-phase deformation model of silica aerogel.
Additionally, the MAP estimate of the mechanical noise multiplier is \blue{0.75 (Table 2)}, showing that for the current model, data, and choices of priors, a smaller total error than the average data variance is needed
to balance between the contribution of data misfit and prior in the Bayesian inference.

\begin{figure}[h!]
	\centering
	\subfloat[]{\includegraphics[width= 0.48\textwidth]{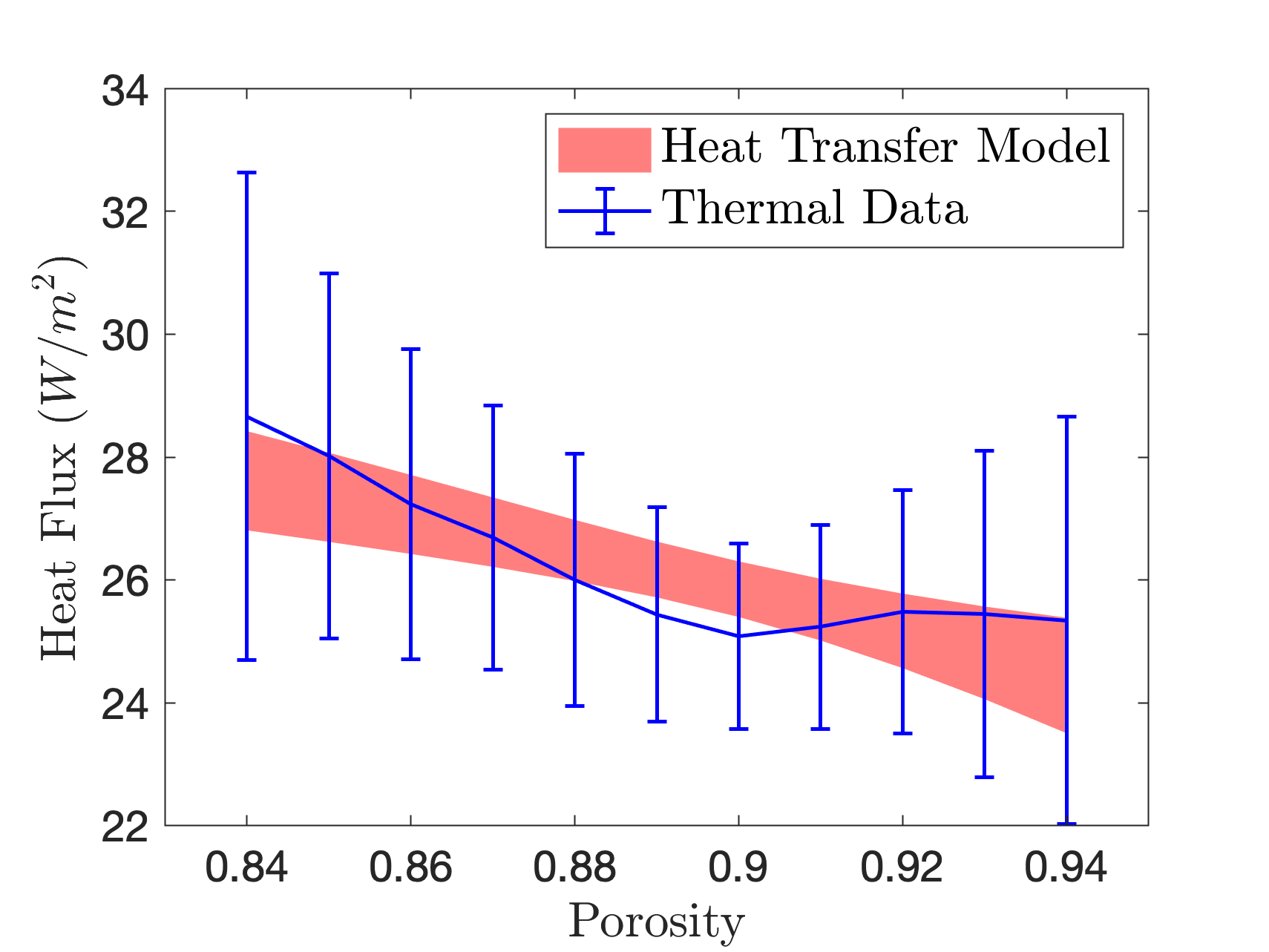}}
	~
	\subfloat[]{\includegraphics[width= 0.48\textwidth]{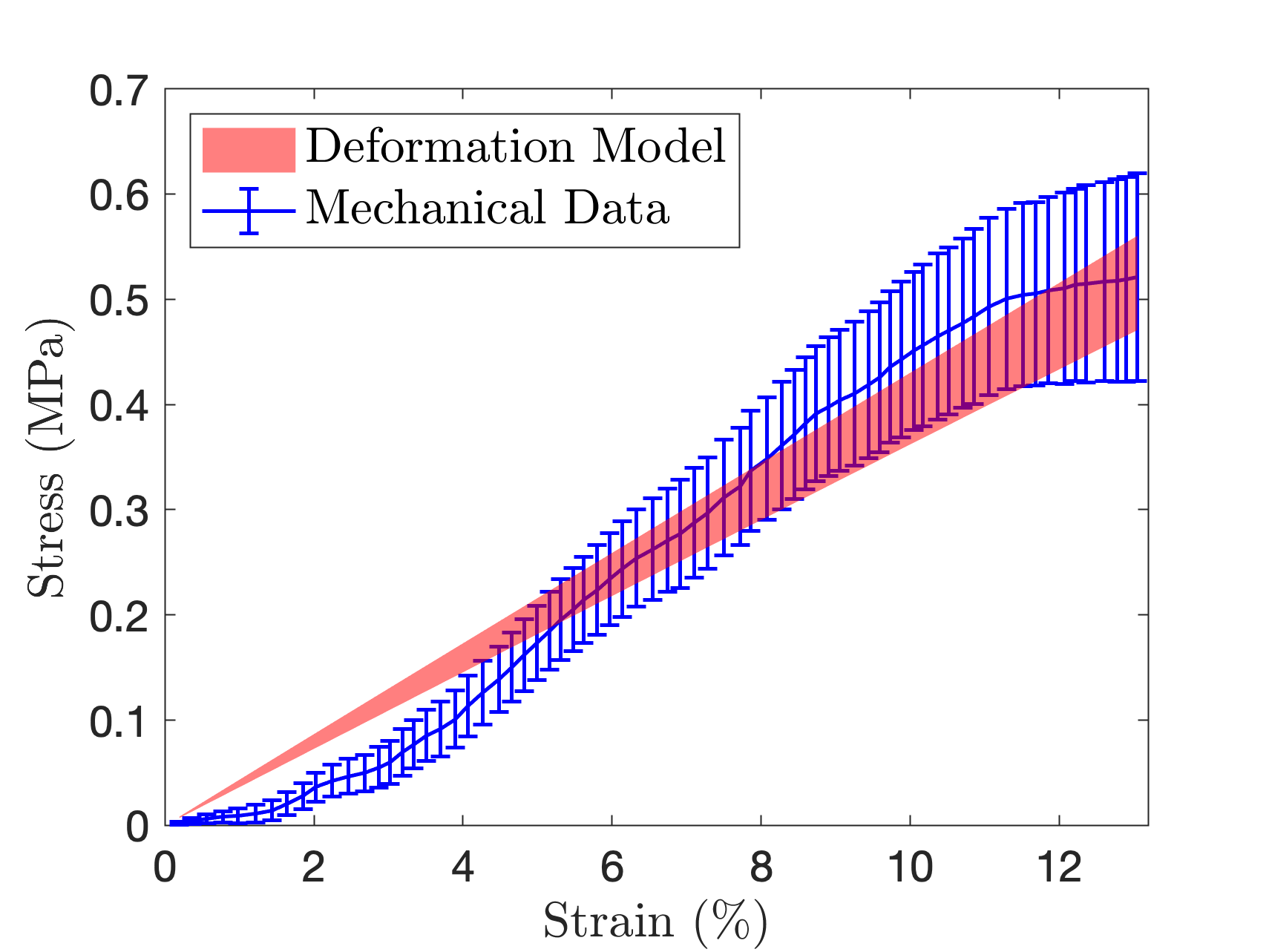}}
	\vspace{-0.1in}
	\caption{
	\blue{
	Comparison of the calibrated models under uncertainty and the experimental data of silica aerogel:
	(a) heat transfer model with an average error of $\epsilon^T = {0.021 \pm 0.732}$; 
	(b) deformation model with an average error of $\epsilon^M = {0.541 \pm 0.524}$. 
	The means and standard deviations for the experimental data are shown in blue, and the red area is the posterior prediction of the model.
	}
	}
	\label{fig:data_model}
\end{figure}

Using the calibrated models with quantified uncertainty characterized by the posterior PDF of the parameters (Figures \ref{fig:thermal_posteriors} and \ref{fig:mechanical_posteriors}), we can solve the statistical forward problem (Section \ref{sec:stat_forward}).
Figure \ref{fig:data_model} compares the heat flux-porosity and stress-strain results computed from the trained models with the corresponding measurement data.
In this figure, the red shaded area shows the posterior prediction of the model, indicating the level of confidence in the model prediction, and the data means and standard deviations are shown in blue. 
The average error in the thermal model, according to the measure \eqref{eq:error}, is $\epsilon^T = \blue{0.021 \pm 0.732}$,
while the error in the deformation model is $\epsilon^M = \blue{0.541 \pm 0.524}$.
The smaller uncertainty in model output compared to the data noise level indicates that the two-phase thermomechanical model of silica aerogel results in variance reduction.
The high error in the deformation model is associated with the simplifying assumptions in the constitutive relations and insufficient data to adequately inform the model parameters. 
The propagation of these uncertainties to the computational predictions of a physical system, in the absence of observational data, is investigated in Section \ref{sec:prediction}.

\begin{figure}[h!]
 \vspace{-0.1in}
	\centering
	\subfloat[]{\includegraphics[width= 0.48\textwidth]{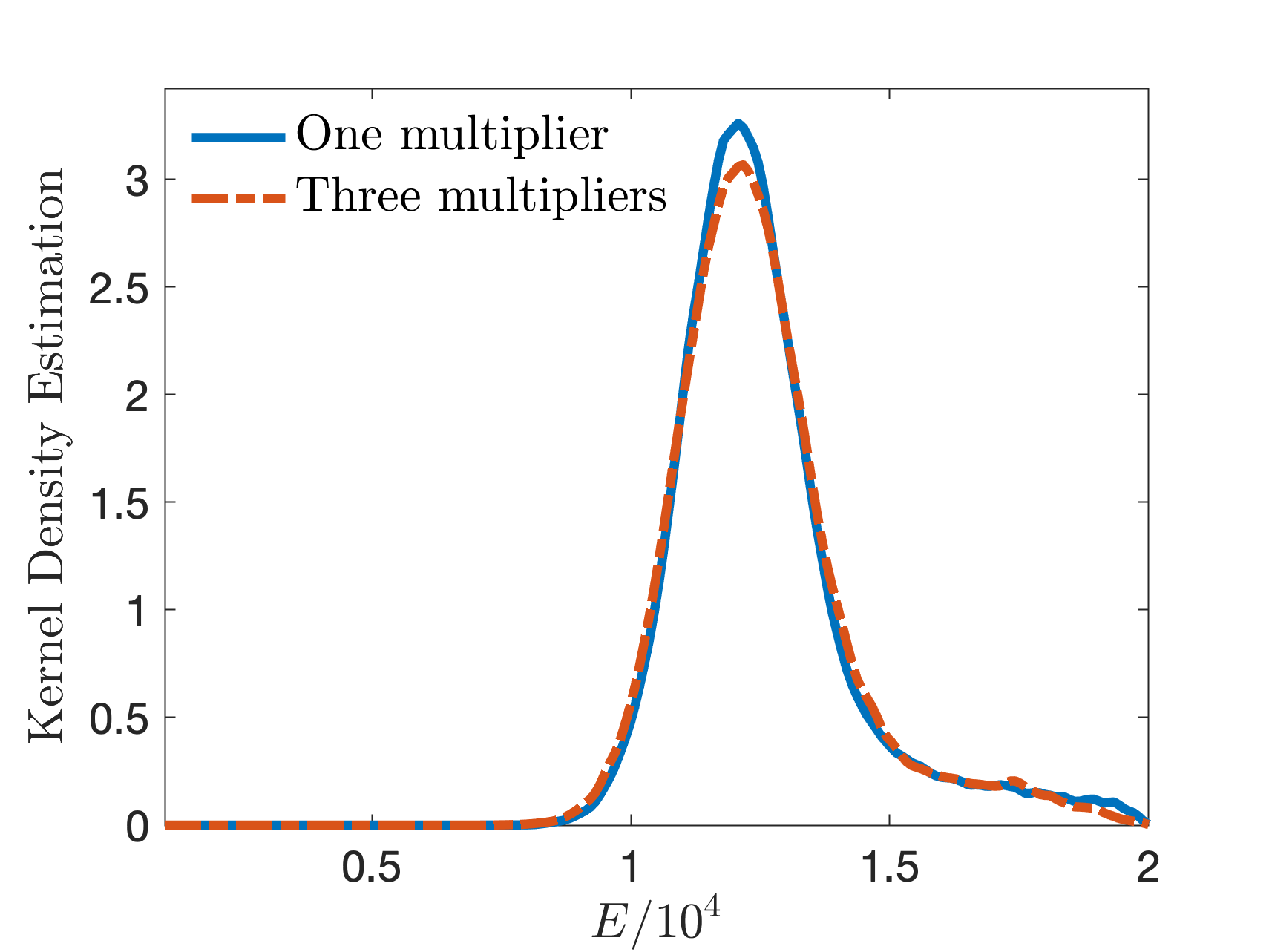}}
	~
	\subfloat[]{\includegraphics[width= 0.48\textwidth]{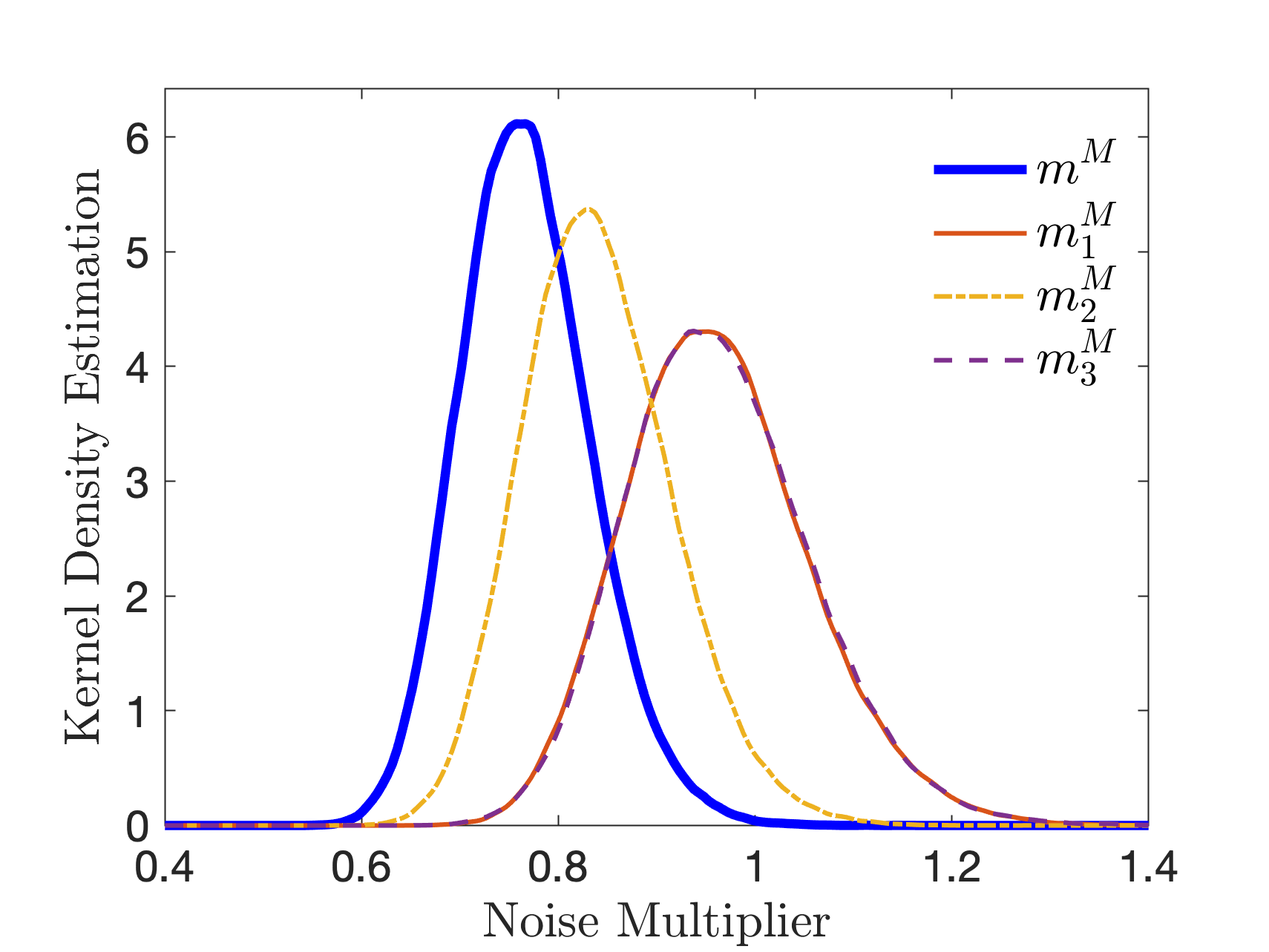}}
	\vspace{-0.1in}
	\caption{
	Comparison of the Bayesian inference results using three noise multipliers and the ones obtained 
	from one noise multiplier (in Figure \ref{fig:mechanical_posteriors}).
	Posterior distributions of the
	(a) the physical parameter $E$; 
	(b) noise multiplier hyper-parameters.
	}
	\label{fig:noise-compare}
\end{figure}

We conclude this section by investigating the impact of the noise model on the Bayesian inference.
As indicated in Section \ref{sec:bayes}, the likelihood function in Bayesian inversion is constructed based upon the uncertainty representation assumptions. 
The choice of noise model presented in \eqref{eq:noise} designates a single hyper-parameter multiplier over the entire data set. 
More accurate learning of the physics-based model from multiple sources of uncertain data may require assigning multiple noise multipliers to different subsets of data.
Nevertheless, increasing the number of multipliers significantly increases the computational costs of Bayesian inversion due to computing the additional posteriors of the hyper-parameters. 
To understand the effect of multiple noise multipliers, the silica aerogel stress-strain measurement data are divided to three subsets for the strain ranges of
$[0 - 3.0 \%]$, $[3.0 \% - 11.0 \%]$, and $[11.0 \% - 12.9 \%]$, consisting of 14, 46, and 12 data points respectively.
The Bayesian inference of the mechanical model is then conducted using the same priors as in {Table 1}, 
while taking into account one noise multiplier for each subset of data.
Figure \ref{fig:noise-compare} shows the posterior distributions of the physical parameter $E$ and the three hyper-parameters compared to the one multiplier inference (Figure \ref{fig:mechanical_posteriors}).
\blue{Table 2} shows the MAP values of the parameter posteriors of the inference using the three multipliers.
For this case, the average error in the posterior prediction of the mechanical model compared to the stress data is $\epsilon^T_{3m} = \blue{0.550 \pm 0.489}$. 
The posterior distributions of the physical parameters and the error between the data and calibrated model are very close for the two choices of noise models, i.e., one and three multipliers.
Figure \ref{fig:noise-compare}(b) shows that despite the different variance levels among the three data subsets, the posteriors of $m_1^M$ and $m_3^M$ are both centered around one (with the MAP estimates of \blue{0.95}). 
On the other hand, the MAP value of posterior of the multiplier $m_2^M$, for the subset with the most data point, is computed as \blue{0.82}.
Similar to the one multiplier case, the smaller than one value of $m_2^M$ indicates that for robust Bayesian inference, the contribution of data misfit should be higher than the prior for the second subset of data.
Comparing the calibrated noise multipliers in Figure \ref{fig:noise-compare}(b) shows 
the number of data points influences the trade-off between the misfit and \blue{uniform} prior in the Bayesian inference.
\begin{table}[h!]
 \label{table:noise}
 \small
 \centering
 \begin{tabular}{c c | c c | c c} 
 \hline
 \hline
 \multicolumn{2}{c|}{\textbf{Heat transfer model}} & \multicolumn{2}{c|}{\textbf{Deformation model}} & \multicolumn{2}{c}{\textbf{Deformation model}} \\
 \hline  
 \textbf{Parameter} & \textbf{MAP} & \textbf{Parameter} & \textbf{MAP} & \textbf{Parameter} & \textbf{MAP} \\ [0.5ex] 
 \hline
 \(\kappa_{s}\) & 0.477 & \(E\) & 11985.0 & \(E\) 		& 11960.1 
 \\
 \(\kappa_{f}\) & 0.085 & \(\log{k}\) & -35.68 & \(\log k\) 	&	-36.14 
 \\
 \(h\) & 81059.0 & \(\log{C}\) & -20.79 & \(\log C\) 	&	-20.72 
 \\  
 \(m^T\) & 0.96 & \(m^M\) & 0.75 & \(m^M_1\) 	&0.95 
 \\   
 &&&& \(m^M_2\) 	&0.82 
 \\  
 &&&&\(m^M_3\) 	 &0.95 
 \\  
 \hline 
 \hline  
 \end{tabular}
 \caption{\blue{
 The Maximum A Posteriori (MAP) estimates of the two-phase heat transfer and deformation model parameters were obtained from the Bayesian inference.
 The first two columns show the inference results considering one noise multiplier for each model, $m^T$ and $m^M$.
 The third column is the calibration results of the deformation model with three noise multipliers, where $m^M_1$, $m^M_2$, and $m^M_3$ are assigned to the data subsets in the strain ranges of $[0 - 3 \%]$, $[3 \% - 11 \%]$, and $[11 \% - 12.9 \%]$, consisting 14, 46, 12 data points respectively.
 }}
\end{table}
%

\subsubsection{Computational prediction under uncertainty}\label{sec:prediction}

\begin{figure}[h!]
 \centering  
 \vspace{-0.1in}
 \includegraphics[trim=0.0in 0in 0in 0.0in, clip, width= 0.55\textwidth]{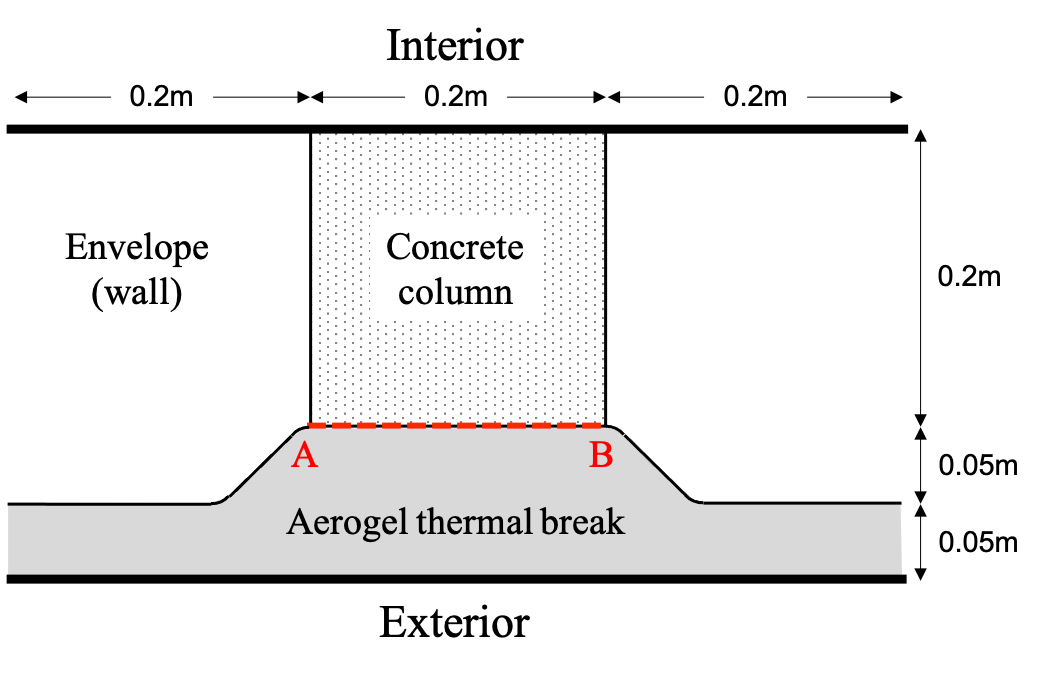}
 \vspace{-0.2in}
 \caption{
 The scenario for the computational prediction. The envelope and column system and the aerogel thermal break alleviate the heat exchange between the exterior and the column.}
 \label{fig:prediction}
\end{figure}

One of the main challenges in building envelope insulation is the energy losses due to thermal bridges at the external elements. Thermal bridging occurs at the interface of envelope components such as wall-roof intersections and beam and pipe penetration to the envelope, leading to heat flow bypassing insulation. Thermal breaks are insulating components incorporated within the building envelope to interrupt the heat flow path and mitigate the effect of thermal bridges, e.g., \cite{goulouti2016}. 
Figure \ref{fig:prediction} shows the horizontal section of a wall and a square concrete column, adopted from \cite{bruggi2011topology}, vulnerable to thermal bridge due to the materials discontinuity.
The column significantly affects the thermal performances of the building envelope since the high conductivity of concrete induces a preferred path for the heat.
A silica aerogel superinsulation thermal break with 90\% porosity is designed to separate the inner column from the building envelope outer boundary to minimize the thermal transmittance. 

\begin{figure}[h!]
 \centering  
 \vspace{-0.1in}
 \subfloat[]{\includegraphics[trim=6.0in 4.5in 6.0in 2.5in, clip, width= 0.48\textwidth]{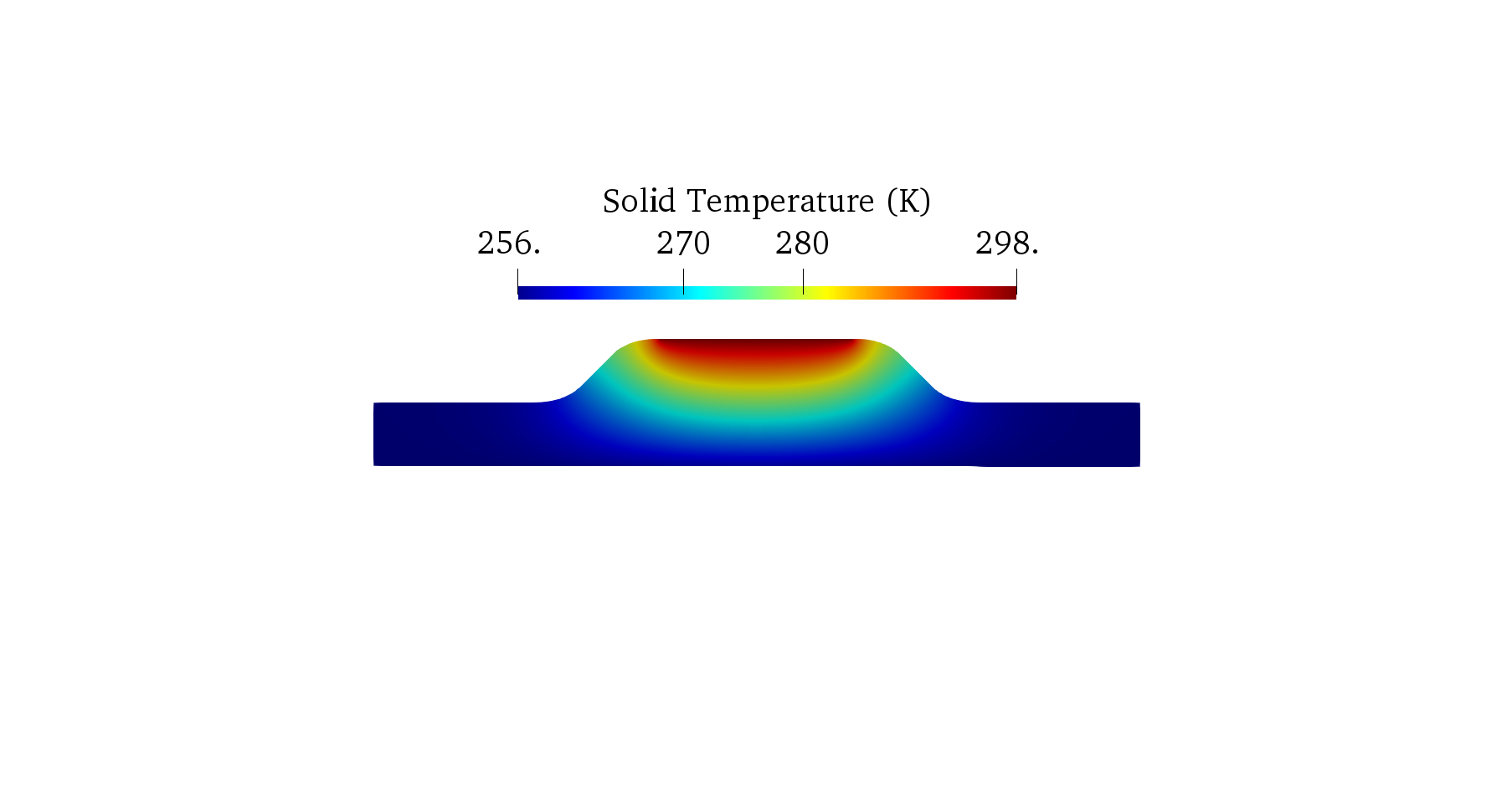}}
 \subfloat[]{\includegraphics[trim=6.0in 4.5in 6.0in 2.5in, clip, width= 0.48\textwidth]{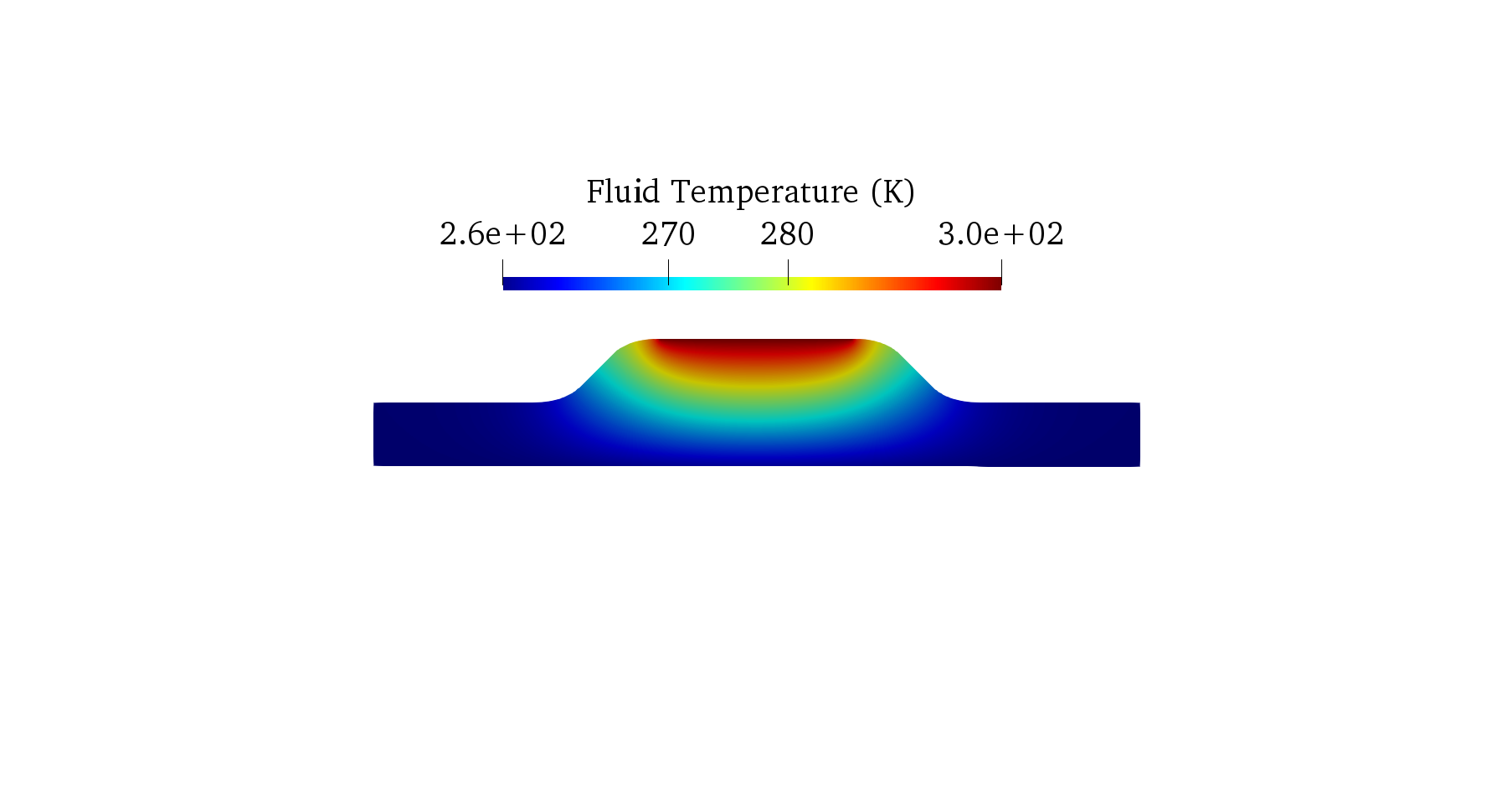}}
	\\ \vspace{-0.1in}
 \subfloat[]{\includegraphics[trim=2.0in 2.5in 2.2in 1.0in, clip, width= 0.48\textwidth]{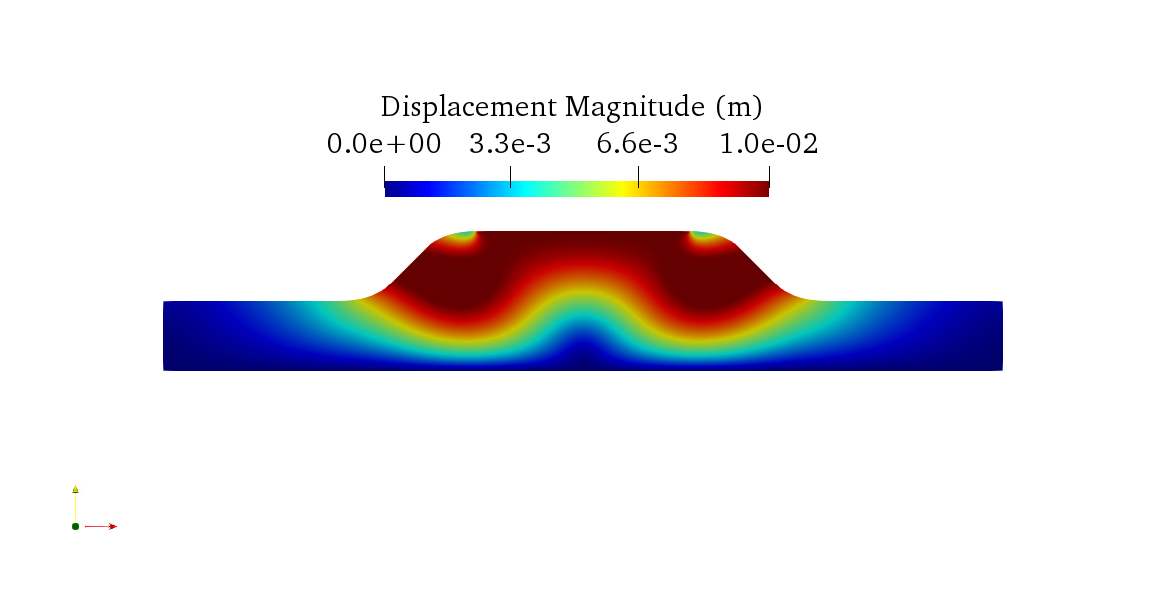}}
 \subfloat[]{\includegraphics[trim=2.0in 2.5in 2.2in 1.0in, clip, width= 0.48\textwidth]{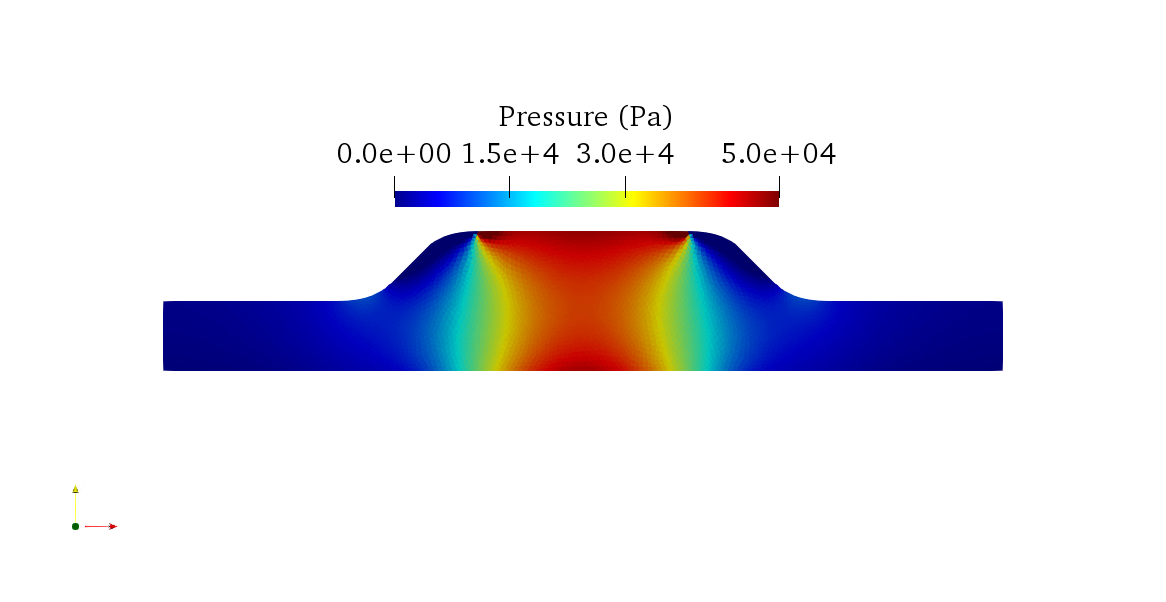}}
 \vspace{-0.1in}
 \caption{
 \blue{Computational prediction of the thermal break responses using the two-phase thermomechanical model 
 and the MAP estimates of the parameters (Table 2). 
 Contour plots of
 (a) steady-state solid temperature $\theta_s$,
 (b) steady-state fluid temperature $\theta_f$,
 (a) magnitude of solid displacement $| \mathbf{u}_s |$ at $t=$100 s,
 (b) fluid partial pressure $p$ at $t=$100 s.}}
 \label{fig:pred_results}
\end{figure}

This section investigates the ability of the calibrated multiphase model with uncertain parameters developed in Section \ref{sec:calibration} to predict thermal and mechanical responses of the aerogel thermal break in Figure \ref{fig:prediction}.
To characterize the thermal insulation of the component, the steady-state heat transfer model \eqref{eq:thermal_bcs} is taken into account together with convective boundary conditions with the ambient temperature $256 K$ (the lowest temperature at Buffalo, NY in 2020) imposed at the exterior boundary and $\bar{\theta}_{AB} = 298 K$ at the interface of the insulation component with the column (surface A-B in the Figure \ref{fig:prediction}).
The resiliency of the thermal break during the construction is investigated by the two-phase deformation model \eqref{eq:mech_bcs}. 
A vertical displacement $\bar{u}_{AB} = 0.01 m$ is applied at the A-B surface of the thermal break over 100 $s$ and at a constant strain rate while the zero displacements are imposed at the bottom.
The fluid pressure is assumed to be zero in all the component's boundaries, representing the fully permeable boundary of the aerogel thermal break.
Figure \ref{fig:pred_results} shows the results of the thermomechanical model prediction using the MAP values of the parameters presented in \blue{Table 2}.

\begin{figure}[h!]
	\centering
	\vspace{-0.3in}
	\subfloat[]{\includegraphics[width= 0.48\textwidth]{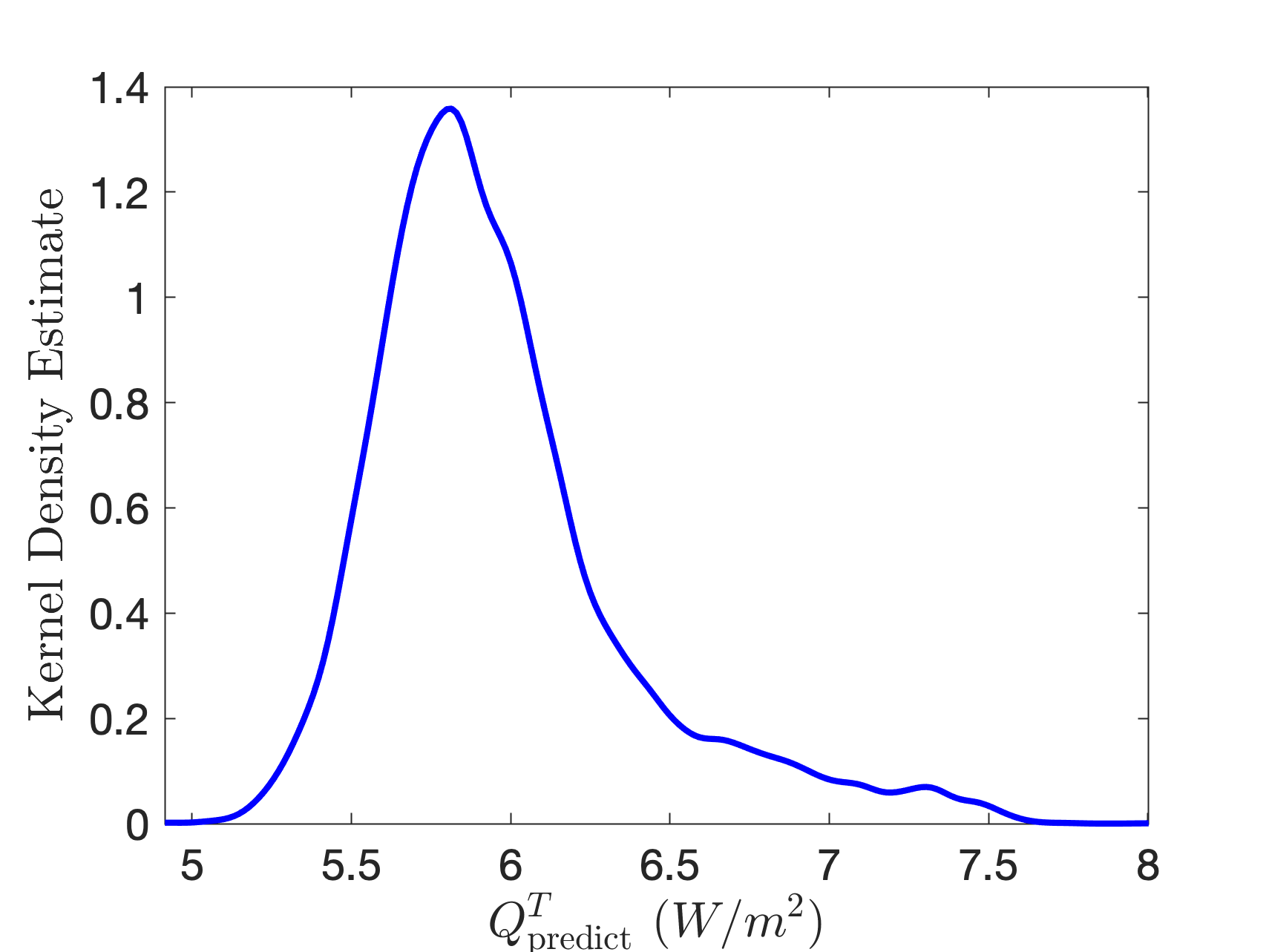}}
	~
	\subfloat[]{\includegraphics[width= 0.48\textwidth]{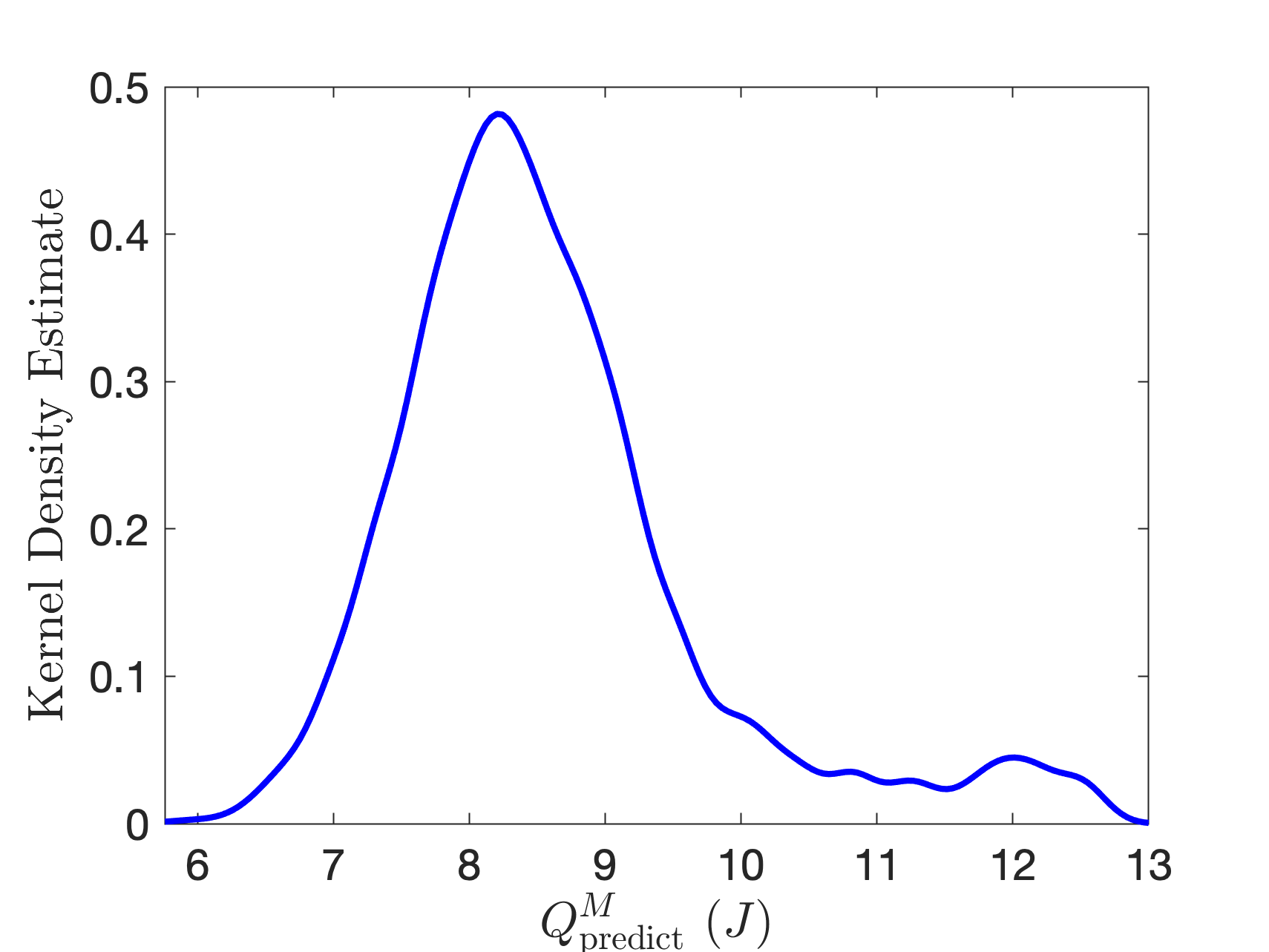}}
	\vspace{-0.1in}
	\caption{
	\blue{
	Probability distributions of the computational predictions of the thermal break in Figure \ref{fig:prediction}:
	(a) thermal QoI as the heat flux of the mixture at the A-B boundary with the mean {5.97} $W/m^2$ and variance {0.17};
	(b) mechanical QoI as the strain energy of the solid phase in \eqref{eq:QoIpredict} with the mean {8.59} $J$ and variance {1.34}.
	}
	}
	\label{fig:QoIpredict}
\end{figure}

Since MAP values are point estimates equivalent to deterministic model calibration, the results in Figure \ref{fig:pred_results} do not provide any information regarding the level of confidence in the computational prediction.
To assess the reliability of the thermomechanical model prediction, we propagate the parametric uncertainty through the model by solving the statistical forward problem in Section \ref{sec:stat_forward}. The target QoIs for the computational prediction of thermal break is the heat flux of the mixture at the A-B boundary in Figure \ref{fig:prediction} (representing how well the component mitigates the undesired thermal bridge) and the strain energy of the solid phase (indicating the resistance of the aerogel \blue{to material damage, i.e., solid wall breakage)},
\begin{eqnarray}\label{eq:QoIpredict}
Q^T_{predict} &=& | q_{AB} |, \nonumber \\
Q^M_{predict} &=& \int_{\Omega} \left(\mathbf{T}_s: \mathbf{E}_s\right) d\mathbf{x}.
\end{eqnarray}
Figure \ref{fig:QoIpredict} shows the probability distributions of the thermal and mechanical QoIs of the thermal break operating in the building envelope, i.e., the prediction scenario obtained from solving the statistical forward problem using \blue{14000} samples of the parameter distributions.
The computational prediction indicates that the heat flux at the A-B boundary is \blue{5.97} $W/m^2$, while the associated parametric uncertainty results in a variance of \blue{0.17}.
Similarly, the strain energy of the silica aerogel is predicted with the mean \blue{8.59} $J$ and variance of \blue{1.34}.
It is seen from these results that while the thermal model prediction is reliable, 
the quantified uncertainty due to modeling assumption and incomplete and noisy data (Section \ref{sec:calibration}) results in significant uncertainty in the deformation model prediction.

\section{Discussion and Conclusions}


\blue{
This work presents a general UQ framework for developing physics-based predictive models of complex materials systems with application to a new multiphase thermomechanical model of porous silica aerogel.
Notable features of this work are the
implementation of a new noise model that characterizes uncertainty in data and modeling error in the Bayesian inference step.
Determining the posterior distributions of the resulting hyperparameters balances the contribution of data misfit in the likelihood function and the parameter priors, resulting in a robust inference process.
The uncertainty analyses are conducted using sampling algorithms, including MCMC for Bayesian inference, 
in which each sample requires the solution of the mixed finite element formulation of the multiphase model. 
To overcome the computational challenge of the UQ framework and ensure efficient use of computing resources, a fully parallelized workload is implemented using open-source libraries for the finite element and the sampling algorithms.
Leveraging the UQ framework, various sources of uncertainty in computational prediction of material behavior are addressed, including:
(i) data uncertainty due to measurements taken from different material specimens,
(ii) microstructural uncertainty due to inadequacy of the model to capture the effect of pore sizes on the material behavior,
(iii) modeling errors incurred by assumptions in the theoretical model (e.g., linear elasticity of the solid phase) and simplifying assumptions in the numerical solution (e.g., considering the 2D model of the inherently 3D physical phenomena).
}

The variance-based global sensitivity analyses of the two-phase model of the silica aerogels with high porosity (around 90\%) indicate that 
\blue{the thermal conductivity of the fluid phase and permeability}
are the most influential model parameters \blue{on the two-phase thermomechanical model outputs for} the silica aerogels with high porosity (around 90\%).
\blue{
The results of the Bayesian inference using experimental data suggests that
the less computationally intensive calibration methods based on the Gaussian assumptions are inadequate for some of the model parameters, leading to inaccurate and unreliable predictions in this class of multiphase models.
The Bayesian model calibration results also show that
despite the significant data uncertainty, due to the microstructural (pore size) variabilities, the two-phase heat transfer model can capture the measurements with approximately 2\% of error, leading to a reliable prediction of thermal performance of the building envelope thermal break. 
However, the deformation model is inadequate to accurately simulate the stress-strain measurements of the silica aerogel, primarily due to the linear isotropic elasticity assumption for the solid phase.
The propagation of uncertainty to the thermal break simulation indicates that
although the average value of the predicted solid strain energy might be below the damage initiation criteria (i.e., solid wall breakage), 
there is significant uncertainty in the mechanical model prediction.
Thus, based on the current model and measurements 
the mechanical resiliency of the aerogel insulation component may be unreliable for building applications. 
}


%
\blue{
There are a few UQ studies of the multiphase behavior of materials in the literature.
For example, Niskanen et al. \cite{niskanen2019} studied the acoustic response of water-saturated ceramic porous materials. The authors conducted Bayesian calibration of a one-dimensional poroelasticity model \cite{biot1955}, posed as first-order ordinary differential equations, and using synthetic measurements.
In comparison, we conducted Bayesian inference using the 2D finite element solution of the two-phase deformation model and  experimental measurements. Despite more computationally intensive analyses, our results depict a more realistic representation of materials' response and associated uncertainty.
In the context of the modeling error assessment within Bayesian inference, Kennedy and O’Hagan \cite{kennedy2001bayesian} presented a general strategy based on assuming errors to be additive to the true physical process.  
Their work motivated various approaches in characterizing the modeling error, including the Bayesian approximation error (BAE) \cite{kaipio2013}, which provides a systematic way to characterize modeling errors based on the availability of a high-fidelity model. 
As another example, Oliver et al. \cite{oliver2015} proposed a framework to embed inadequacy representations in the model components for which we have low confidence (e.g., constitutive relations) while respecting high confidence components (e.g., conservation and balance laws). Such inadequacy representation is formulated from the existing knowledge regarding the system and is calibrated and validated using data.
Additionally, following \cite{kennedy2001bayesian}, Zhang et al. \cite{ZHANG201915} considered emulating the physics-based model output and the associated modeling errors with Gaussian processes. Utilizing such surrogate models speeds up the Bayesian solutions to obtain the parameter posteriors.
In comparison, our method addresses modeling errors by calibrating noise hyperparameters while solving the Bayesian inference problem.
While the approaches presented by \cite{kaipio2013} and \cite{oliver2015} will lead to a more accurate assessment of modeling error in a particular problem, our approach is nonintrusive and applicable to a broad range of physical systems. This is because our general approach neither requires a validated high-fidelity model as in BAE nor the problem-dependent inadequacy representations as in \cite{oliver2015}.
Additionally, although the sampling algorithm to compute the parameter posteriors using the finite element solution of the model is computationally expensive compared to \cite{ZHANG201915}, our method ensures preserving the physical laws governing the material model (see section \ref{sec:mixture}). Pure data-driven surrogate models such as Gaussian processes disregards the predictive power of the physics-based models.
}

Despite the comprehensive uncertainty analyses of the two-phase thermomechanical model silica aerogel in this work, several areas can be addressed in future studies. 
Additional experimental studies are required to characterize the underlying mechanisms of the newly fabricated silica aerogel and guide enhancing the multiphase mixture model. For instance, the stress-strain measurements, including loading and unloading, indentation experiments, e.g., \cite{faghihi2012indent, CHEN2016}
\blue{, and microstructure imaging before and after compression testing} 
can provide more insights on the \blue{micro-mechanical responses and the effect of fluid pressure on the silica aerogel deformation.}
Additionally, a higher fidelity simulation of the hierarchical porous structure and pore size on the silica aerogel properties may be achieved by gradient-enhanced nonlocal models, such as those in \cite{faghihi2014jemt, faghihi2012thermal}. This class of models possesses length and time scales parameters to account for the microstructural interactions on the macroscopic behavior of the aerogel.
Moreover, various computational models with different fidelity and complexity can be constructed to simulate aerogel behavior. Given sufficient experimental data, implementation of the Bayesian model validation framework, based on the Occam-Plausibility Algorithm \cite{farrell2015jcp, odenbabuska2017}, enables adaptive selection of the optimal model for predicting the silica aerogel responses.
Finally, the reliable prediction of 3D printed aerogel components, e.g., thermal breaks, must cope with the spatial uncertainty due to the layer-by-layer bonding in additive manufacturing.

%
%

In summary, this study shows that making a reliable computational prediction of complex materials systems beyond observation requires taking advantage of physics-based models and accurate characterization of uncertainties in modeling and measurements with the aid of high-performance computing.

\section*{Acknowledgments}
We benefited from the discussions with Dr. Kathryn Maupin, of the Sandia National Laboratories, on the software applications and the Bayesian inference solution.

\bibliographystyle{abbrv} 
\bibliography{refs}
\index{Bibliography@\emph{Bibliography}}%

\end{document}